\documentclass[12pt]{article}

\usepackage{graphicx}
\usepackage{epstopdf}
\usepackage{amsmath}
\usepackage{amssymb}
\usepackage{amsthm}
\usepackage{ccaption}
\usepackage[usenames]{color}
\usepackage[
      colorlinks=true,
      linkcolor=blue,  
      urlcolor=blue,   
      citecolor=blue, 
      filecolor=blue,     
      linkcolor=blue,
      pdfstartview=FitV,
      pdftitle={},
        pdfauthor={Mukund Rangamani, Simon Ross},
        pdfsubject={},
        pdfkeywords={tachyons, btz},
        pdfpagemode=None,
        bookmarksopen=true    
      ]{hyperref}

\makeatletter
\@addtoreset{equation}{section}

\makeatletter
\renewcommand\section{\@startsection {section}{1}{\z@}%
                                   {-3.5ex \@plus -1ex \@minus -.2ex}
                                   {2.3ex \@plus.2ex}%
                                   {\normalfont\large\bfseries}}
\renewcommand\subsection{\@startsection{subsection}{2}{\z@}%
                                     {-3.25ex\@plus -1ex \@minus -.2ex}%
                                     {1.5ex \@plus .2ex}%
                                     {\normalfont\bfseries}}

  \captionnamefont{\bfseries}
  \captiontitlefont{\small\sffamily}
  \captiondelim{: }
  \hangcaption

\def\baselinestretch{1.2}
\newcommand{\be}{\begin{equation}}
\newcommand{\ee}{\end{equation}}
\newcommand{\beq}{\begin{eqnarray}}
\newcommand{\eeq}{\end{eqnarray}}

\def\sec#1{\S \ref{#1}}

\def\req#1{(\ref{#1})}
\def\App#1{Appendix \ref{#1}}

\def\[{\left [}
\def\]{\right ]}
\def\({\left (}
\def\){\right )}

\def\cf{{\it cf.}}
\def\ie{{\it i.e.}}

\def\a{\alpha}
\def\b{\beta}

\def\d{\delta}
\def\eps{\epsilon}

\def\vth{\vartheta}

\def\p{\partial}

\def\om{\omega}
\def\lam{\lambda}

\def\CH{{\cal H}}

\def\CL{{\cal L}}

\def\CN{{\cal N}}
\def\CO{{\cal O}}

\def\CX{{\cal X}}
\def\CY{{\cal Y}}
\def\CZ{{\cal Z}}
\def\ZZ{\mathbb{Z}}

\def\cC{\mathbb{C}}

\def\R{{\bf R}}
\def\Sp{{\bf S}}

\def\A5S5{{\rm AdS}_5 \times \S^5}

\def\Id{1\!\!1}
\def\inv{^{-1}}
\def\l{\ell}
\def\half{{\frac{1}{2}}}

\def\p{\partial}

\def\half{{\frac{1}{2}}}

\def\p{\partial}

\def\Tr#1{{\rm Tr}\(#1\)}

\def\ket#1{\mid  \! #1  \rangle}

\newcommand{\bbibitem}[1]{\bibitem{#1}\marginpar{#1}}

\def\Label#1{\label{#1}%
{ \color{blue}{\smash{\hbox to0pt{\raise2ex\hbox{\tiny[#1]}\hss}}}}}
\def\noLabels{\let\Label=\label}
\def\nobbibitem{\let\bbibitem=\bibitem}

\def\ads3{{\rm AdS}$_3$}
\def\adss3{{\rm AdS}$_3$ $\times$ $\Sp^3$}
\def\slt{$SL(2,\R)$}
\def\slc{$\widehat{SL(2,\R)}$}
\def\slcur#1{$\widehat{SL(2,\R)_{#1}}$}
\def\sucur#1{$\widehat{SU(2)_{#1}}$}

\def\comm#1#2{{[#1,#2]}}

\def\tk{\tilde{k}}
\def\nonu{\nonumber \\}

\def\etau{\tilde{\tau}}

\title{{\bf \Large Winding tachyons in BTZ}}

\author{\normalsize 
Mukund Rangamani\footnote{mukund.rangamani@durham.ac.uk} \;and
Simon F. Ross\footnote{s.f.ross@durham.ac.uk} \\ \\ 
\small \sl Centre for Particle Theory \& Department of
Mathematical Sciences,
\\[-1.5mm]
\small \sl Durham Univerity, South Road, Durham DH1 3LE, United Kingdom. \\
}

\begin{document}

\noLabels 
\nobbibitem 

\setlength{\baselineskip}{16pt}
\begin{titlepage}
\maketitle
\begin{picture}(0,0)(0,0)
\put(350, 303){DCPT-07/21} 
\end{picture}
\vspace{-36pt}

\begin{abstract}
  Analysing closed string tachyon condensation will improve our
  understanding of spacetime in string theory. We study the string
  spectrum on a BTZ black hole spacetime supported by NS-NS flux,
  which provides a calculable example where we would expect to find a
  quasi-localised tachyon. We find that there is a winding tachyon
  when the black hole horizon is smaller than the string scale, as
  expected. However, due to effects of the NS-NS $B$ field, this
  tachyon is not localised in the region where the spatial circle is
  string scale. We also discuss the relation to the Milne orbifold in
  the limit near the singularity.
 \end{abstract}
\thispagestyle{empty}
\setcounter{page}{0}
\end{titlepage}

\renewcommand{\baselinestretch}{1.4}  
\renewcommand{\thefootnote}{\arabic{footnote}}



\section{Introduction}
\label{intro}

The study of tachyons in string theory provides an interesting window
into aspects of non-perturbative dynamics. The dynamics of open string
tachyon condensation is relatively well understood, but the
corresponding story for closed string tachyons is far from complete.
We have a nice picture for localised closed string tachyons at
orbifold singularities such as $\cC/\mathbb{Z}_N$ \cite{Adams:2001sv}
(\cf, \cite{Headrick:2004hz} for a review), where by virtue of the
tachyon dynamics being confined to a small region in spacetime, one
has control over the condensation process. Recently, there has been
interest in studying quasi-localised closed string tachyons
\cite{Adams:2005rb}, which have been argued to arise in several
interesting contexts. The basic idea is that for a string on a circle
of size  smaller than the string length $\l_s$,  with anti-periodic boundary condition for
fermions, there are tachyonic winding modes. If the size of this
circle varies over some base space, one heuristically expects a
tachyon which is confined to the region where the size of the circle
$\le \l_s$.  Such configurations arise when we consider strings
propagating on a Riemann surface in corners of moduli space where
handles degenerate \cite{Adams:2005rb}, in simple time-dependent
spaces \cite{McGreevy:2005ci}, or in charged black string geometries
\cite{Horowitz:2005vp, Ross:2005ms}.  The condensation of such
tachyons is argued to provide insight into issues such as spacetime
fragmentation/topology change, black hole evaporation, and spacelike
singularity resolution
\cite{McGreevy:2005ci,Horowitz:2006mr,Silverstein:2006tm}. (In the
last context, the tachyon condensate provides a realisation of the
final state proposal of \cite{Horowitz:2003he}.)

Most of the discussion of quasi-localised tachyons so far has been
based on this kind of approximate analysis, as the examples considered
were too complicated for the string spectrum to be calculated
explicitly. In this paper, we consider in detail the string spectrum
on a BTZ black hole ($\times \Sp^3 \times {\bf
  T}^4$)~\cite{Banados:1992wn,Banados:1992gq}. The arguments used
previously imply that the BTZ black hole has a winding tachyon when
the horizon size $\sqrt{k} r_+ \le \l_s$ \cite{Horowitz:2005vp}, and
that this tachyon will be confined to the region near the horizon,
where the spatial circle is smaller than the string scale. Indeed,
this geometry arises as the near-horizon limit of the black string
examples considered in \cite{Horowitz:2005vp, Ross:2005ms}.

In BTZ, we can calculate the perturbative string spectrum exactly, and
test this heuristic analysis. The BTZ black hole is an orbifold of
\ads3 by an identification under a boost. We consider the AdS$_3
\times \Sp^3 \times {\bf T}^4$ geometry supported by NS-NS flux,
corresponding to the F1-NS5 system in Type II string theory
compactified on\footnote{We can alternately consider compactification
  on K3. The internal space will play no role in our analysis, and we
  will concentrate on ${\bf T}^4$ for simplicity.}  ${\bf T}^4$.  The
world-sheet theory is a CFT with a \slcur{k} $\times$ \sucur{k}
super-current algebra, with the level $k$ being set by the NS-NS flux, or
alternatively by the number of effective strings in six-dimensions.
The bosonic string on the BTZ orbifold has been previously studied
in~\cite{Natsuume:1996ij,Satoh:1997xe,Hemming:2001we,Martinec:2002xq,Hemming:2002kd}.
We exploit and extend these results to determine when there is a
winding string tachyon in the BTZ geometry.

We find that there is indeed a twisted sector tachyon in the spectrum,
which for the superstring appears precisely when $\sqrt{k} r_+ \le
\sqrt{2}\l_s$. In the superstring, the tachyon in odd twisted sectors
will survive the GSO projection if the spin structure on spacetime
imposes anti-periodic boundary conditions on fermions around the
spatial circle~\cite{Scherk:1978ta}.  This is in accord with the expectations from the
qualitative argument. 

The major surprise of our analysis is that the
tachyon wavefunctions are not localised!  We find that the tachyon has
non-trivial support all the way out to the AdS boundary, with a
wavefunction very similar to that for a bulk tachyon.  The NS-NS flux
plays a key role in this delocalisation. It is directly related to the
existence of `long string' states in this geometry, which can grow
arbitrarily large due to the cancellation of the string tension by the
coupling to the background $B$ field~\cite{Seiberg:1999xz}. This
delocalisation will make it more difficult to understand the
condensation of these tachyons. However, one might hope that the AdS
asymptotics might result in  the tachyon condensation only appreciably
changing the geometry in some compact region. 

We also study the Milne limit, where we zoom in on the region near the
singularity. This limit is analogous to the flat space limit of the
elliptic orbifolds of~\cite{Martinec:2001cf}. We find that with an
appropriate scaling, physical states survive in both twisted and
untwisted sectors in the limit. We argue that from the T-dual point of
view, these twisted sectors seem to be localised near the singularity,
in agreement with the expectations of~\cite{McGreevy:2005ci}. We leave
a detailed understanding of the relation of the twisted sectors we find
here to previous work on the Milne
orbifold~\cite{Cornalba:2002fi,Nekrasov:2002kf,Pioline:2003bs} for future investigation.

In the next section, we briefly outline the relevant aspects of string
theory on \ads3 and the BTZ black hole. We then discuss the
computation of the twisted sector tachyon for the bosonic string in
\sec{bstring}, and for the superstring in \sec{sstring}.  We conclude
with some remarks on open issues in \sec{discuss}. Our conventions for
\slt\ are contained in \App{ads3btz}. We review the flat-space limit
of the elliptic orbifold in \App{ellorb}. We briefly discuss aspects
of the thermal AdS partition function in \App{partitionfn}.

{\bf Note added:} After this paper was completed, we learnt that
tachyons in BTZ have also been investigated from a Euclidean
perspective in~\cite{Lin:2007gi,Micha}.

\section{Preliminaries}
\label{background}

To set the stage for discussing string theory on the BTZ background,
we collect some useful information regarding the the WZW model with
target space \ads3 and the \slc\ current algebra. Further details
regarding our conventions can be found in \App{ads3btz}.

\subsection{AdS$_3$}

Bosonic string theory on \ads3 with NS-NS flux is described by an
\slt\ WZW model (see e.g.~\cite{Maldacena:2000hw} for a nice
discussion). The action for the WZW model is the conventional one
\begin{equation}
S_{WZW} ={k \over 8 \pi\,\a'} \,\int \, d^2 \sigma \, \Tr{g^{-1} \,\p_a g \, g^{-1}\,\p^a g} + {i k\over 12 \pi} \, \int \,\Tr{ g^{-1}dg \wedge  g^{-1}dg \wedge  g^{-1}dg}. 
\Label{wzwact}
\end{equation}	
The level $k$ of the WZW model is not quantised, since $H^3(\slt, \R) =
0$. Later, when we discuss the superstring, we will quantise $k$, since
the level of the \slt\ current algebra will be tied to that of an
$SU(2)$ current algebra (for strings on \adss3). For purposes of
discussing the \ads3 geometry, the \slt\ group manifold is
conveniently parametrised in terms of global coordinates $(t, \rho,
\phi)$ as\footnote{This choice corresponds to the Euler angle
  parametrisation of $SU(1,1)$. The isomorphism between \slt\ and
  $SU(1,1)$ given by $g\in SL(2,\R) \implies h = t\inv g t \in SU(1,1)
  $ where $t = \Id + i\, \sigma_1$. }
\begin{equation}
 g =  \left(\begin{array}{cc} \cos \tau \, \cosh \!\rho + \sin
      \theta \, \sinh \!\rho & \sin \tau \, \cosh \!\rho + \cos \theta \,\sinh \!\rho \\
      - \sin \tau \,\cosh \!\rho + \cos \theta \,\sinh \!\rho & \cos \tau\,\cosh \!\rho -
      \sin \theta \, \sinh\! \rho \end{array}\right),
\Label{globalp}
\end{equation}
which leads to the metric
\begin{equation}
ds^2 = \a' k\, \(-\cosh^2\!\!\rho \, d\tau^2 + d \rho^2 + \sinh^2\! \!\rho
\,   d\theta^2 \) 
\Label{globalmet}
\end{equation}	
and NS-NS two-form
\begin{equation}
B = \alpha' k \,\sinh^2\!\!\rho \,d\tau \wedge d\theta.
\end{equation}
Henceforth, we will set $\alpha' = 1$, so we work in units of the
string length. The AdS length scale is then $\ell = \sqrt{k}$. 

The WZW model \req{wzwact} is invariant under the action
\begin{equation}
g(z, \bar{z} ) \to \om(z) \, g(z,\bar{z})\, \bar{\om}(\bar{z})^{-1} ,
\Label{wzwinv}
\end{equation}	
which leads to a set of conserved world-sheet currents\footnote{We are
  using the $\tau^a$ generators for \slt; see the appendix for our
  conventions.} 
\begin{equation}
J^a = k \, \Tr{\tau^a \, \p g \, g^{-1}}.
\Label{wzwcurrents}
\end{equation}	
This choice of currents ensures that in the flat space limit $k\to
\infty$, $J^a$ reduce to the translational currents. The conformal
Ward identity implies the OPEs
\begin{equation}
J^a(z) \, J^b(w) \sim {k\over 2} \, {\eta^{ab} \over (z-w)^2} +
{i\eps^{ab}_{\ \ c} \, J^c(w)\over (z-w)}, 
\Label{jjope}
\end{equation}	
with a similar expression for the right-movers.\footnote{Our
  conventions for the \slc\ are analogous to those used in
 \cite{Martinec:2001cf}. As discussed there we need to redefine the
  right-moving currents to ensure that the standard conventions for
  raising and lowering operators is respected. We assume henceforth
  that the appropriate redefinition has been applied to the
  right-movers.} The OPE can be translated into commutation relations
by using the mode expansions
\begin{equation}
J^a(z) = \sum_{n = -\infty}^{\infty} \, J_n^a \, z^{-n-1},
\Label{curmode}
\end{equation}	
leading to 
\begin{eqnarray}
\comm{J_n^3}{J_m^3} &=& -{k \over 2} \, n\, \d_{n+m,0},  \nonu
\comm{J_n^3}{J_m^\pm} & =& \pm \, J_{n+m}^\pm, \nonu
\comm{J_n^+}{J_m^-} & =& -2\, J^3_{n+m} + k\, n\, \d_{n+m,0}.
\Label{comrelcur}
\end{eqnarray}	
Here we have used $J^\pm = J^1 \pm i J^2$. This choice corresponds to
the elliptic basis of \slt\ used for \ads3 or spacelike quotients
thereof~\cite{Martinec:2001cf}, and is useful if we want to
diagonalise $J^3(z)$.

The world-sheet Virasoro generators are 
\begin{eqnarray}
L_0 &=& {1\over k-2} \, \[\(J_0^1\)^2 + \(J_0^2\)^2
 - \(J^3_0\)^2 + 2\, \sum_{m=1}^{\infty}  
\, \( J_m^1 J_m^1 + J_m^2 J_m^2 - J_m^3 J_m^3\) \] ,\cr
L_{n\neq0} & = & {2\over k-2}\, \sum_{m=1}^{\infty} 
\, \( J_{n-m}^1 J_m^1 + J_{n-m}^2 J_m^2 - J_{n-m}^3 J_m^3\),
\Label{virgen}
\end{eqnarray}
with commutation relations:
\begin{equation}
\comm{L_n}{L_m} = (n-m) \, L_{n+m} + {c \over 12}\, n (n^2 -1) \, \d_{n+m,0}
\Label{vcom}
\end{equation}	
and
\begin{equation}
[L_n,J^a_m] = -m J^a_{n+m}.
\Label{curvcom}
\end{equation}	
The central charge $c$ is given in terms of the level $k$ as 
\begin{equation}
c = {3 k \over k-2} \ . 
\Label{csltwo}
\end{equation}	
Note that the contribution to $L_0$ from the zero modes of the currents is proportional to the quadratic Casimir $c_2$ of \slt.

The spectrum of strings on global \ads3 contains the untwisted, or
short string states in the representations of the current algebra
$\hat {\mathcal C}_j^\alpha \times \hat {\mathcal C}_j^\alpha$, $j =
\frac{1}{2} + is$ and $\hat {\mathcal D}_j^\pm \times \hat {\mathcal
  D}_j^\pm$ for $ \frac{1}{2} < j < \frac{k-1}{2}$. These current
algebra representations are highest weight representations of the
current algebra built from the corresponding \slt\ representations by
acting with current algebra lowering operators. The $\mathcal
C_j^\alpha$ are continuous representations of \slt, while $\mathcal
D_j^\pm$ are respectively highest and lowest weight discrete series
representations. The continuous representations correspond to the
bosonic string tachyon; this follows from the fact that the quadratic Casimir 
is  $-j(j-1)$.  The spectrum on global \ads3 will also
contain twisted sector states obtained by acting on these short string
states with spectral flow, as described in~\cite{Maldacena:2000hw}.
In~\cite{Argurio:2000tb}, it was shown that this spectral flow could
be re-expressed in terms of twisting with respect to a twist operator
which imposes the periodicity in global coordinates. In our case, we
will have instead twisted sectors corresponding to the BTZ orbifold.

\subsection{BTZ}

We will study the non-rotating BTZ black hole,\footnote{This is a
  simpler example since the action of the orbifold is left-right
  symmetric. The generalisation to the rotating case involves an
  asymmetric orbifold.} which is an orbifold of \ads3 by a hyperbolic
generator of \slt~\cite{Banados:1992gq}. To describe this orbifold, we
use a different parametrisation of the group.  Describing the AdS
space in BTZ coordinates amounts to writing the \slt\ group element in
Euler angles~\cite{Natsuume:1996ij}:
\begin{equation}
g  = e^{-2 \, i\,\varphi' \tau^3} e^{-2 \, i \, \rho' \tau^1} e^{-2 \, i\, \psi' \tau^3} = \left(\begin{array}{cc}e^{\varphi'} & 0 \\ 0   &
    e^{-\varphi'} \end{array}\right)\left(\begin{array}{cc}r  &
    \sqrt{r^2-1} \\ \sqrt{r^2-1}  &
    r \end{array}\right)\left(\begin{array}{cc}e^{\psi'} & 0
      \\ 0   & e^{-\psi'} \end{array}\right),
\Label{btzeul}
\end{equation}	
where $r = \cosh \!\rho'$. In these coordinates, the target space
metric of the WZW model \req{wzwact} is:
\begin{equation}
ds^2 = k \left[ - (r^2-1) dt^2 +
    \frac{dr^2}{r^2-1} + r^2 d\phi^2 \right] ,
\Label{btzmet}
\end{equation}	
where $\phi = (\varphi' + \psi')$, $t = (\varphi'-\psi')$. The background  NS-NS two-form can be
written in a suitable gauge as
\begin{equation}
B = k \,(r^2-1) \,d\phi \wedge dt  \ .
\Label{bfldbtz}
\end{equation}	
The orbifold action which generates a non-rotating BTZ black hole is
then simply $\phi \sim \phi + 2\pi r_+$. Note that $r_+$ is
dimensionless and $M_{BH} = r_+^2$. Unlike \eqref{globalp}, the
coordinates in \req{btzeul} do not cover the full spacetime; they are
valid outside the event horizon $r=1$, where the proper size of the
$\phi$ circle is $2\pi \sqrt{k}\, r_+$.

This choice of basis for the generators can now be translated into the
current algebra. The BTZ coordinates correspond to choosing a
hyperbolic basis for the current algebra, in which the generator $J^2$
is diagonalised, as the generators of
spacetime time translation and rotation are~\cite{Hemming:2001we}
\begin{equation}
Q_t = J_0^2 - \bar J_0^2, \quad Q_\phi = J_0^2 + \bar
  J_0^2.
\Label{gens}
\end{equation}	
Since these involve $J^2_0$, we are interested in real eigenvalues of
$J_0^2$.  The commutation relations for the current algebra in the
hyperbolic basis read 
\begin{eqnarray}
\comm{J_n^2}{J_m^2} &=& {k \over 2} \, n\, \d_{n+m,0},  \nonu
\comm{J_n^2}{J_m^\pm} & =& \pm i\, J_{n+m}^\pm, \nonu
\comm{J_n^+}{J_m^-} & =& 2i\, J^2_{n+m} + k\, n\, \d_{n+m,0},
\Label{hypcomrelcur}
\end{eqnarray}	
where we have used $J^\pm = J^1 \pm J^3$. Note that $J_m^\pm$ have
$J_0^2$ charge $\pm i$. The issues associated with this are
discussed in detail in\footnote{See~\cite{Kuriyan:1968lr} for an excellent discussion of the representations in the hyperbolic basis.}~\cite{Natsuume:1996ij,Hemming:2001we}. The corresponding OPEs are (\cf \req{jjope} )
\begin{eqnarray}
J^+(z) J^-(w) &\sim& \frac{k}{(z-w)^2} + \frac{2i J^2}{(z-w)}\ , \nonu
J^2(z) J^2(w) &\sim& \frac{k/2}{(z-w)^2}, \nonu
  J^2(z) J^\pm(w) &\sim& \pm \frac{i J^\pm}{(z-w)}.
\Label{hypope}
\end{eqnarray}	
It will also be useful for later discussion to record the explicit
form of the currents in the BTZ coordinates.  In the parametrisation
\req{btzeul} we find that the currents \req{wzwcurrents} take the form
\begin{equation}
J^1 = ik \,\(\cosh 2\varphi'\, \partial \rho' - 2\sinh 2 \varphi'
\, \cosh \!\rho' \sinh \!\rho'  \,\partial \psi'\) \ ,
\Label{btzcurj1}
\end{equation}
\begin{equation}
J^3 = ik \(\sinh 2 \varphi' \, \partial \rho' 
- 2\cosh 2 \varphi' \, \cosh\! \rho' \sinh \!\rho'  \, \partial \psi'\), 
\Label{btzcurj3}
\end{equation}
\begin{equation}
J^2 = ik \,\(\partial \varphi' + (\cosh^2\!\! \rho' + \sinh^2 \!\!\rho')
\partial \psi'\) ,
\Label{btzcurj2}
\end{equation}
where we write $r = \cosh \!\rho'$. Similarly, the anti-holomorphic
currents are written as
\begin{equation}
\bar J^1 = ik \,\(\cosh 2\psi' \, \bar \partial \rho' - 2\sinh 2 \psi'
\cosh \!\rho' \sinh \!\rho'  \,\bar \partial \varphi'\),
\Label{btzcurjb1}
\end{equation}
\begin{equation}
\bar J^3 = ik \,\(-\sinh 2 \psi' \,\bar \partial
\rho' + 2\cosh 2 \psi' \cosh\! \rho' \sinh\! \rho'  \,\bar \partial \varphi'\), 
\Label{btzcurjb3}
\end{equation}
\begin{equation}
\bar J^2 = ik\, \(\bar \partial \psi' + (\cosh^2\!\! \rho' + \sinh^2 \!\!\rho')
\bar \partial \varphi'\) . 
\Label{btzcurjb2}
\end{equation}

Bosonic strings in the BTZ background were originally studied in
~\cite{Natsuume:1996ij,Satoh:1997xe} and more recently in
\cite{Hemming:2001we}.  The latter analysis reproduced the spectrum by
applying the spectral flow operation introduced in
~\cite{Maldacena:2000hw} to generate the twisted sectors. Our aim is
to more explicitly identify the tachyon in these twisted sectors. We
will also extend the analysis of the orbifold to the superstring.

\section{The bosonic string}
\label{bstring}

As we have seen above, the BTZ black hole is obtained by a quotient of
\slc\ by a hyperbolic element. In the BTZ coordinates \eqref{btzeul},
the quotient is simply the identification $\phi \sim \phi+ 2\pi\, r_+$. We
want to understand the twisted sectors associated with this orbifold,
and see under what circumstances we will find a tachyon in the twisted
sectors.

\subsection{Twisted sectors of the BTZ orbifold}
\label{twistedbtz}

The periodic identification along $\partial_\phi$ which generates the BTZ
orbifold restricts the states to have quantised values of $Q_\phi$. By
\eqref{gens}, this restricts the $J_0^2 + \bar{J}_0^2$ eigenvalue:
\begin{equation}
r_+  \(J_0^2 + \bar{J}_0^2\) \in \ZZ \ , 
\Label{orbrestrict}
\end{equation}	
where $J_0^2$ refers to the eigenvalue of the corresponding operator
on the states. In addition to this restriction on the untwisted
sectors, the orbifold action will introduce appropriate twisted
sectors. Following~\cite{Argurio:2000tb}, we find it convenient to
determine the twisted sectors by imposing the
constraint~\eqref{orbrestrict} on an enlarged set of vertex operators. 
We implement this by first introducing an appropriate twist
operator $t_n$, and then projecting onto the states which are mutually
local with respect to this twist operator. The twisted sector vertex
operators are then obtained by taking the set of operators including
the twist operator which are mutually local and closed under OPE.

To construct twisted sectors, it is convenient to work with a
parafermionic representation of the current algebra (analogous to the
construction of \cite{Martinec:2001cf} in the elliptic
case).\footnote{This choice of representation is inspired by the
  analysis of \cite{Martinec:2001cf}, where the orbifolds
  \ads3$/\ZZ_N$ involving identifications of \ads3 (and extensions to
  include the orbifold also acting on the internal CFT) under the
  spatial rotation isometry $\partial_\theta$ were studied.
  In fact the parafermion OPEs written in \req{paraope} are the same
  as in the parafermionic representation of the elliptic form of
  \slcur{k}. In that case the $J^3$ current is bosonised in terms of a
  free field; see \App{ellorb} for some details.} To begin with we bosonise the $J^2$ current in terms of
a free field $X$;
\begin{equation}
J^2 = -i \sqrt{\frac{k}{2}} \partial X\ ,
\Label{hyppara1}
\end{equation}	
where $X(z)\,X (w)\sim - \ln (z-w)$, and introduce parafermions to represent
the remaining \slcur{k}$/ \widehat{U(1)}$ algebra by
\begin{equation}
J^\pm = \xi^\pm e^{\pm \sqrt{\frac{2}{k}} X} \ ,
\Label{hyppara2}
\end{equation}	
with
\begin{equation}
\xi^+ \xi^- \sim \frac{k}{(z-w)^{2 + \frac{2}{k}}}, \quad  \xi^\pm \xi^\pm \sim (z-w)^{\frac{2}{k}} \ .
\Label{paraope}
\end{equation}	
For chiral
primary operators of the current algebra, there is a parafermionic
representation
\begin{equation}
\Phi_{j\lambda}(w) = \Psi_{j\lambda}(w) e^{-i \sqrt{\frac{2}{k}}
  \lambda X}\ ,
\Label{chiralpara}
\end{equation}	
where $\lam$ is the $J^2$ eigenvalue, which determines the spacetime energy. Note that in the hyperbolic basis $\lam$ and  $j$ are unrelated. The primary operators have conformal dimension 
\begin{equation}
h(\Phi_{j\lam}) = -{j(j-1)\over k-2}
\Label{primdim}
\end{equation}	
where $c_2 = -j(j-1)$ is the Casimir of the global \slt\ symmetry generated by the zero modes of the currents. For the continuous representations $c_2 \ge {1\over 4}$; it is bounded from above, $c_2 \le {1\over 4}$, for the  discrete representations. Non-tachyonic modes are required to have  $c_2 \le{1\over 4}$ which corresponds to the Breitenlohner-Freedman bound in \ads3. From \req{chiralpara} and \req{primdim} it  follows that
\begin{equation}
h(\Psi_{j\lambda}) = - \frac{j(j-1)}{(k-2)} - \frac{\lambda^2}{k}\ .
\Label{paradim}
\end{equation}	
In this parafermionic representation, the restriction
\req{orbrestrict} can be imposed by introducing twist operators
\begin{equation}
t_n = e^{i r_+ \sqrt{\frac{k}{2}} n (X- \bar X)} \qquad {\rm for} \; n \in \ZZ \ ,
\Label{twist}
\end{equation}	
and requiring that physical vertex operators are mutually local with
respect to these twist operators.

Given the twist operator it is easy to write down the vertex operators
for primary states in the $n^{{\rm th}}$ twisted sector. They are just
given by the composite operator arising from the product of the
untwisted sector primary with the twist, \ie,
\begin{equation}
\Phi^n_{j\lambda \bar \lambda} = \Psi_{j \lambda} \bar
  \Psi_{j \bar \lambda} e^{-i \sqrt{\frac{2}{k}} [(\lambda +
    \frac{k}{2} n r_+) X + (\bar \lambda - \frac{k}{2} n r_+) \bar
    X]},
\Label{twistst}
\end{equation}	
where $\Psi_{j \lambda}$, $\Psi_{j \bar \lambda}$ are the chiral
parafermions from the untwisted sector primaries. These operators have
dimensions 
\begin{eqnarray}
h(\Phi^n_{j\lam{\bar \lam}})  &=& -\frac{j(j-1)}{(k-2)} - \frac{\lambda^2}{k} +
  \frac{(\lambda + k r_+ n/2)^2}{k} = -\frac{j(j-1)}{(k-2)} + \lambda
  r_+ n + \frac{k n^2 r_+^2}{4}, \cr
   \bar h(\Phi^n_{j\lam{\bar \lam}}) &=& -\frac{j(j-1)}{(k-2)} -
  \frac{\bar \lambda^2}{k} + \frac{(\bar \lambda - k r_+ n/2)^2}{k} =
  -\frac{j(j-1)}{(k-2)} - \bar \lambda r_+ n + \frac{k n^2 r_+^2}{4}.
\Label{dims}
\end{eqnarray}	

In~\cite{Hemming:2001we}, these twisted sectors were discussed using
the language of spectral flow developed in~\cite{Maldacena:2000hw}.
For global AdS, the spectral flow is equivalent to the introduction of
an appropriate twist operator, as discussed in~\cite{Argurio:2000tb}.
However, for the BTZ orbifold, we think the twist operator language is
more appropriate, as the twisting does not correspond to an
automorphism of the full current algebra. The symmetries associated
with $J^\pm$ are broken by the orbifold ($J^\pm$ are not mutually
local with respect to $t_n$), so these operators will have different
moding in the twisted sectors. This twisting is still related to a
spectral flow: if we focus on the algebra of the surviving symmetries,
which is the $\widehat{U(1)}$ algebra generated by $J^2$ and the
Virasoro algebra, the spectral flow
\begin{equation}
\tilde J^2_n = J^2_n + \frac{k}{2} w
  \delta_{n,0} \ ,  \qquad \tilde L_n = L_n + w J^2_n + \frac{k}{4} w^2
  \delta_{n,0} 
\Label{spectralflow}
\end{equation}	
for arbitrary $w$ is an automorphism of this algebra. Taking $w = n
r_+$, $\bar w = -n r_+$ for integer $n$ recovers the charges of the
twisted sector states described above. However, this restricted
algebra is no longer spectrum generating. 

The full vertex operators are formed by taking descendants of the
primary operators \eqref{twistst} and combining them with some vertex
operator from the internal CFT. The physical state conditions $(L_0 -1) \ket{{\rm phys}} = (\bar{L}_0 -1) \ket{{\rm phys}} =0$ will
then be
\begin{equation} \label{physl}
-\frac{j(j-1)}{(k-2)} - \frac{\lambda^2}{k} +
  \frac{(\lambda + k r_+ n/2)^2}{k} + h_{int} + N = 1,
\end{equation}
\begin{equation} \label{physr}
 -\frac{j(j-1)}{(k-2)}  -
  \frac{\bar \lambda^2}{k} + \frac{(\bar \lambda - k r_+ n/2)^2}{k} +
 \bar h_{int} + \bar N = 1,
\end{equation}
where $h_{int}, \bar h_{int}$ are the dimensions of the operator from
the internal CFT, and $N, \bar N$ are oscillator numbers for the current algebra. We assume that the internal CFT is unitary, so $h_{int}, \bar h_{int} \geq 0$. 

Finally, we should consider the relation of $\lambda, \bar \lambda$ to
spacetime energy more carefully. It is clear that $J^2_0 + \bar J^2_0$
corresponds to momentum around the compact circle, but there are two
possible contributions to $J^2_0 - \bar J^2_0$, coming from spacetime
energy or winding around the compact circle. That is, there is an
ambiguity in the definition of $Q_t$ in the twisted sectors, analogous
to the ambiguity in the definition of $Q_\phi$ discussed
in~\cite{Hemming:2001we}. If we apply the naive formula \eqref{gens},
the twisted sector operators have energy
\begin{equation} \label{qt1}
E = \lambda - \bar \lambda + k r_+ n, 
\end{equation}
since the eigenvalue of $J^2_0$ is $\lambda + kr_+ n/2$ and the
eigenvalue of $\bar J^2_0$ is $\bar \lambda - k r_+ n/2$, for a twisted
sector vertex operator \eqref{twistst}. However, thinking of our
orbifold as analogous to an ordinary translation orbifold to generate
a compact circle, this twist contribution to the $J^2_0, \bar J^2_0$
eigenvalue is more naturally interpreted as the usual winding
contribution to $p^L_\phi$, $p^R_\phi$. Therefore we do not think it
is appropriate to interpret it as a contribution to the spacetime
energy of the mode. We therefore propose to identify instead
\begin{equation} \label{qt2}
Q_t = J^2_0 - \bar J^2_0 - k r_+ n
\end{equation}
as the generator of spacetime time translation, so that the spacetime
energy of the mode \eqref{twistst} is simply $\lambda -\bar
\lambda$. As explained in~\cite{Hemming:2001we}, this shift
corresponds to adding the divergence of an antisymmetric tensor to the
Noether current; this does not change the conservation law, but shifts
the value of the charge in topologically nontrivial sectors. 

This issue becomes clearer when we study the flat space limit. In
\sec{btzflatnh}, we will see that \eqref{qt2} gives the usual
notion of spacetime energy in the translational orbifold. It should be 
noted that the appropriate choice is actually  gauge dependent. 
We will return to this issue  in \sec{milne} where \eqref{gens} is a more 
appropriate choice of generators in the chosen gauge.

\subsection{Tachyons in BTZ}
\label{btztachyons}

Having determined the spectrum of twisted sector operators in the BTZ
orbifold, we want to determine which of them corresponds to a tachyon
in the spacetime. We first need to consider carefully the question of
how a tachyon is defined. A mode is tachyonic if it has sufficiently
negative spacetime mass-squared. We want to apply this condition by
thinking of our orbifold as analogous to a translational orbifold, and
looking for modes which have appropriately negative
mass-squared\footnote{As we are dealing with an asymptotically AdS
  geometry, the appropriate condition for a tachyon is that the mass
  squared violates the Breitenlohner-Freedman (BF) bound, which for
  \ads3 is $m^2 \le -{1\over4}$. } in the directions orthogonal to the
orbifold.

We are twisting with respect to $J^2_0$, so we
view the Casimir
\begin{equation} \label{2dcas}
J^1_0 J^1_0 - J^3_0 J^3_0 = \frac{1}{2}( J^+_0 J^-_0 + J^-_0 J^+_0) 
\end{equation}
for the other two components of the current as representing the
directions orthogonal to the orbifold.  Note that although $J^\pm_0$
individually do not commute with $J^2_0$, this Casimir will, so we can
work with a basis of vertex operators which are eigenvectors for this
Casimir. In the parafermionic representation, the eigenvalue of this
Casimir is a multiple of the dimension of the parafermionic part of
the vertex operator \eqref{paradim}, so what we want to do is to view
the parafermionic part of the operator as representing the
contribution from the orthogonal dimensions. This is not strictly true
in a naive sense, since the bosonic field $X$ introduced to bosonise
$J^2$ is not simply a target space coordinate on the circle.
Nonetheless, we think this is a natural interpretation. We would then
decompose \eqref{dims} into the dimension of the parafermionic
operator, \eqref{paradim}, and a contribution
\begin{equation}
\frac{(\lambda + k r_+ n/2)^2}{k} \Label{hcirc}
\end{equation}
associated with the compact circle.

For general operators, there is a problem, as this latter term depends
on the spacetime energy $Q_t$ as well as the momentum $Q_\phi$ on the
compact circle. This dependence on $Q_t$ is a complicating factor, so
we will focus for now on identifying tachyon operators with $Q_t = 0$,
that is, $\lambda = \bar \lambda$. If there is a field with mass
squared violating the BF bound, it will have a mode with zero energy,
so this analysis should still be sufficiently general to find all
spacetime tachyons, at least in the region outside the horizon. In
this case, $\lambda = Q_\phi/2$, and we can interpret \eqref{hcirc} as
$p_L^2$, the usual contribution of the momentum and winding on a
compact circle to the conformal dimension.  Thus in this case, an
appropriate criterion to identify a tachyon is that the Casimir of the
representation in the space orthogonal to the orbifold direction
should be $\ge {1\over4}$.  That is, we claim that the appropriate
criterion for a twisted or untwisted sector mode with $\lambda = \bar
\lambda$ to be tachyonic is that the parafermionic part of the
operator has positive dimension greater than ${1\over 4\,(k-2)}$.

We see that unlike in the case of the elliptic orbifolds analysed
in~\cite{Martinec:2001cf}, we can only get tachyons from operators in
the continuous representations, even when we are considering the
twisted sectors. For \eqref{paradim} to be greater than ${1\over
  4\,(k-2)}$, we need the full quadratic Casimir $-j(j-1)$ to violate
the BF bound.  The discrete representations of \slt\ at best saturate
the bound. The essential difference between the elliptic and
hyperbolic cases is the sign of the second term in \eqref{paradim}.

We want to construct physical states which are tachyonic. The
dimensions of operators in the internal CFT will be positive, so to be
able to satisfy the physical state condition, we need to require in
addition that the total dimensions of the \slt\ vertex operator
\eqref{twistst} are $h, \bar h \leq 1$.\footnote{In the more familiar
  case of orbifolding in the internal CFT, a tachyon is also
  identified with a relevant operator, but the argument is different:
  there, the dimension of operators in the CFT which includes the time
  direction could be negative, but we require it to be positive to
  have a tachyon, and therefore need $h \leq 1$ for the internal CFT.
  Here, $h$ is the dimension of an operator in the BTZ CFT, which
  includes the time direction, so we need $h \leq 1$ to be able to
  satisfy the physical state condition for any choice of operator in
  the internal CFT. Note however that not any relevant operator in
  this BTZ CFT corresponds to a tachyon: only those which satisfy the
  additional condition that \eqref{2dcas} is sufficiently positive
  do.}  With our restriction to $\lambda=\bar \lambda$, this condition
is most easily satisfied for zero momentum, $\lambda = \bar \lambda
=0$, when
\begin{equation}
h = \bar{h } = -\frac{j(j-1)}{(k-2)} + \frac{k n^2 r_+^2}{4} = \frac{{1\over4}+s^2}{k-2} + \frac{k n^2 r_+^2}{4} \ ,
\Label{diml0}
\end{equation}	
where we have used the $j$ value for a principal continuous
representation, $j = \half + is $. The condition $h \leq 1$ thus
translates (for large $k$) to $\sqrt{k}r_+ < 2$. Thus, we conclude
that there will be tachyons in the twisted sectors if and only if
$\sqrt{k}r_+ < 2$. The vertex operator corresponding to the most
tachyonic mode is $\Phi_{j00}^n$ with $j = \frac{1}{2} + is$. Note
that in the contrary case $\sqrt{k} r_+ > 2$, we see no tachyon in the
spectrum for $\lambda = \bar \lambda$. 

The bound $\sqrt{k}r_+ < 2$ is in good agreement with what we expect
based on the heuristic argument comparing this space to a
Scherk-Schwarz compactification. In the next subsection, we will study
the near-horizon limit, and recover the usual Scherk-Schwarz analysis~\cite{Scherk:1978ta} as a limit of the present discussion.

\subsection{Flat space limit of BTZ}
\label{btzflatnh}

There are two interesting flat space limits which we can consider by
sending the AdS curvature to zero. Firstly, we can zoom in on the
near-horizon region keeping the part of the spacetime outside the
horizon, and secondly we zoom in on the singularity. For the moment we
will concentrate on the first case and return to the second later. In
this limit, the generator we are orbifolding along goes over to a
translation generator in flat space, and our orbifold reduces to the
usual Scherk-Schwarz compactification. 

In the first limit, we need to take $k \to \infty$ holding the horizon
radius in AdS units $R = \sqrt{k} r_+$ fixed. Let us define
coordinates
\begin{equation}
x^2 = \sqrt{k} \phi \ , \;\;  \rho = \sqrt{k}
\sqrt{r^2-1} = \sqrt{k} \sinh \rho' \ ,
\Label{nhcoord}
\end{equation}	
in which the metric becomes:
\begin{equation}
ds^2 = -\rho^2 dt^2 + d\rho^2 + (dx^2)^2 + \CO\({1\over k}\) .
\Label{nhmetr}
\end{equation}
Note that $x^2$ is a periodic coordinate, $x^2 \sim x^2 + 2\pi R$. The
metric \req{nhmetr} is just two dimensional Rindler times a circle.
Further defining coordinates $x^1 = \rho \cosh t$, $x^3 = \rho \sinh
t$, the metric becomes
\begin{equation}
ds^2 = -(dx^3)^2 + (dx^1)^2 + (dx^2)^2,
\end{equation}
The currents are to leading order simply $J^a = i \sqrt{k}\, 
\partial x^a$, $\bar J^a = i \sqrt{k} \, \bar \partial x^a$ which are
translational currents in the flat metric.

However, to understand the time translation and momentum generators in
the near-horizon region, we need to be more careful, and keep track of
sub-leading terms in $J^2$, $\bar J^2$. Recall that the rotation
generator $Q_\phi = J^2_0 + \bar J^2_0$; hence $p_2$ will have a
finite value in the near-horizon limit if $\lambda + \bar \lambda \sim
\sqrt{k}$.  On the other hand, the energy is $E
= \lambda - \bar \lambda$, so it is finite if $\lambda - \bar \lambda \sim 1$.
We therefore need to consider the terms in $J^2$ which are $\CO(1)$ to
see the $t$-translation generator. Retaining terms to sub-leading
order, we find
\begin{equation}
J^2 = i \sqrt{k}\, \partial x^2 - i \rho^2 \, \partial t, 
\Label{jfleft}
\end{equation}
\begin{equation}
\bar J^2 = i \sqrt{k} \, {\bar \partial} x^2 + i \rho^2 \,{\bar \partial} t. 
\Label{jfright}
\end{equation}
Thus in this flat space limit, 
\begin{equation}
J^2 - \bar J^2 = i \sqrt{k}\, (\partial - {\bar \partial}) x^2 - i
\rho^2 \, (\partial + {\bar \partial})t,
\end{equation}
and we can see quite clearly that there are two contributions, one
$\CO(\sqrt{k})$ associated with winding, and one $\CO(1)$ associated
with time translation. This shows why we need to take a winding part
out of $J^2_0 - \bar J^2_0$ to obtain $Q_t$ in \eqref{qt2}. 

It might seem surprising that these currents \req{jfleft} and
\req{jfright} are conserved holomorphic and anti-holomorphic currents;
in flat space, the Lorentz invariance only implies
\begin{equation}
{\bar \partial}\(\rho^2\, \partial t\) + \partial \(\rho^2 \, {\bar
  \partial} t\) 
= 0,
\Label{lorcons}
\end{equation}
not separate conservation of the left- and right-moving parts. In
fact, it is the total $J^2$ which is conserved, not each term
separately. To see why the currents \req{jfleft} and \req{jfright} are
conserved, we need to work with the equations of motion to sub-leading
order, including a term coming from the $B$ field.  In the
near-horizon limit, it is convenient to work with the $B$ field in the
gauge \req{bfldbtz}. In the near-horizon limit we then have a
$B$-field
\begin{equation}
B = \frac{1}{\sqrt{k}} \rho^2 dx^2 \wedge dt. 
\Label{bnhor}
\end{equation}
This makes a sub-leading contribution to the $x^2$ equation of motion
\begin{equation}
\partial \bar \partial x^2 + \frac{1}{2 \sqrt{k}} \(\partial(\rho^2
\bar \partial t) - \bar \partial (\rho^2 \partial t)\) = 0 .
\end{equation}
Together with the conservation law following from Lorentz invariance
\req{lorcons}, this indeed implies the conservation of $J^2$, $\bar
J^2$ to the indicated order.

Now, it is clear that in this flat space limit, a tachyon is a mode
which has a negative mass-squared in the subspace spanned by $x^3,
x^1$. That is, if we consider a vertex operator of zero momentum in
the $x^2$ direction, with winding $n$, and write the conformal
dimension as
\begin{equation} \label{flatt}
h = \bar h = C + \frac{n^2 R^2}{4},
\end{equation}
then the operator is a tachyon if $C$ is positive,\footnote{Of course, in taking the flat space limit we are no longer sensitive to the finite $k$ piece coming from the BF bound. The criterion espoused in  \sec{btztachyons}, $h(\Psi_{j\lam}) \ge {1\over 4 \,(k-2)} $, simply reduces to the positivity of the Casimir in the two dimensions.} as this is the
Casimir in the $x^3, x^1$ directions. In AdS$_3$, if we start with an
untwisted sector operator with $\lambda = \bar \lambda = 0$, and apply
$n$ units of twist, the conformal dimension of the resulting twisted
sector state is
\begin{equation} \label{flatt2}
h = \bar h = - \frac{j (j-1)}{(k-2)} + \frac{k n^2 r_+^2}{4}. 
\end{equation}
Comparing \eqref{flatt} to \eqref{flatt2}, we see that the state
corresponds to a tachyon in the twisted sector if and only if it comes
from a tachyon -- a continuous representation -- in the untwisted
sector, precisely as we argued in the previous section. Thus, we see
that in this near-horizon limit, the space is approximately flat, with
one direction periodically identified, and the twisted sector tachyons
identified in the previous section go over precisely to the usual
Scherk-Schwarz winding tachyons in the flat space. This shows how the
approximate Scherk-Schwarz analysis can be recovered from our exact analysis.

\subsection{(Non)localisation of tachyon}
\label{nonlocalt}

One of our main aims is to say something about the localisation of
this winding tachyon. It is difficult to analyse this precisely, as we
need to understand the spacetime dependence of the twisted sector
vertex operators. We have seen in the previous section that the
tachyons all come from operators in the continuous representations of
\slt. In~\cite{Natsuume:1996ij}, the radial profile of the 
vertex operator wavefunction for untwisted sectors was analysed in
terms of hypergeometric functions. From this analysis, we can see that
as expected, the untwisted sector tachyon of the bosonic string is not
localised in the radial direction. 

It is not completely straightforward to extend this analysis to the
twisted sectors, as the twisted sector vertex operators $\Phi^n_{j00}$
differ from the untwisted vertex operator by a phase factor $e^{-i
  \,\frac{\sqrt{k}}{2}\,r_+n \(X-\bar{X}\)}$, and the field $X$ is not
simply related to the target space coordinates. However, using the
definition of $X$ \req{hyppara1} and the currents in BTZ
\req{btzcurj2},\req{btzcurjb2}, we can see that $\p X \propto \(r^2
\,\p\phi - (r^2 -1)\,\p t\)$, so we would expect that there is no
exponential damping with the radial direction $r$ coming from the
twist field. So the radial profile of the wavefunction is roughly the
same as the untwisted vertex operator. As a result, it appears that
the twisted sector tachyons are also not localised!

This conclusion can be further supported and understood by considering
the analysis in the T-dual description of the CFT. The winding mode
then becomes an ordinary momentum mode, and the analysis in the T-dual
geometry can be performed at a supergravity level. Note however that
in the full geometry the $\phi$ circle has a size determined by the
radial coordinate $r$ and therefore the T-dual has a varying dilaton
that becomes strongly coupled deep inside the bulk. This would
invalidate working with tree level string theory. Nonetheless, this
T-dual analysis provides some indication of the behaviour
of the vertex operator wavefunctions, and gives some more intuitive
understanding of the failure of the mode to be localised.
See~\cite{Dijkgraaf:1991ba} for a related discussion in the context of
the two dimensional black hole.

The T-dual of the BTZ black hole was worked out
in~\cite{Horowitz:1993jc}. The geometry is
\begin{equation} \Label{tdual}
ds^2 = -\frac{k (r^2-1)}{r^2} dt^2 + \frac{2}{r^2}(r^2 - 1) dt
d\theta + \frac{d\theta^2}{r^2 k} + \frac{k
  dr^2}{(r^2-1)}, 
\end{equation}
the dilaton is 
\begin{equation}
e^{-2\phi} =  k r_+^2 r^2,
\end{equation}
and the $B$ field vanishes in this T-dual description. The coordinate
$\theta$ parametrises the T-dual circle, and has periodic
identifications $\theta \sim \theta + 2\pi/r_+$. The determinant of
the metric is $g = -1/r^2$, and the inverse metric is
\begin{equation}
g^{-1} = \left( \begin{array}{ccc} -\frac{1}{k (r^2-1)} & 1 & 0
    \\ 1 & k & 0 \\ 0 & 0 & \frac{(r^2-1)}{k} \end{array} \right). 
\end{equation}

We want to consider a mode with one unit of momentum on $\theta$,
which is T-dual to the first winding mode. As a warm-up, we can
consider the geodesics. The geodesic equation reduces to
\begin{equation}
\dot r^2 - E^2 = \frac{(r^2-r_+^2)}{k} (-m^2 - k L^2  + 2 k
E L),
\end{equation}
where $E,L$ are the conserved quantities associated to $\partial_t,
\partial_\theta$, and $m$ is the particle's rest mass. We can see that
the effect of the angular momentum is to effectively shift the
mass-squared by a finite amount; in particular, the effect is
independent of radius. The $r$ dependence comes solely from
red-shifting of the radial momentum.
Considering the wave equation for a scalar field $T$ of mass $m$,
if we set $T = f(r) e^{i \omega t} e^{iL \theta}$, we have
\begin{equation} \label{twave}
  r \partial_r \left( \frac{(r^2- 1)}{k r} \partial_r f \right)
  + \left( \frac{\omega^2}{k (r^2-1)} - 2 \omega L - k L^2
  \right) f = m^2 f,
\end{equation}
and again the angular momentum acts just as a shift on the effective
mass. In both cases, the essential point is that the inverse metric
component $g^{\theta \theta} = k$, so the contribution of this
momentum is independent of radius. Since $L = n r_+$ for integer $n$,
this is precisely reproducing the contribution from the winding modes
in the original description. If we consider a mode with $\omega =0$,
the effective mass $\tilde m^2 = m^2 + k L^2$ corresponds to the mass
of the mode in a Kaluza-Klein reduced 2d theory. Hence, the tachyonic
modes are those for which $\tilde m^2 <\tilde m^2_{BF}$, and they behave in
exactly the same way for $L=0$ and $L \neq 0$: the winding tachyons
have the same radial wavefunction as a non-winding tachyon with the
same value of $\tilde m^2$. Hence, our winding tachyons are not
localised in the near-horizon region.

This T-dual analysis makes it clear that the failure of the tachyon to
be localised is due to the coupling to the $B$ field in the original
spacetime. If we considered a BTZ geometry with no $B$ field (for
example, the S-dual D1-D5 geometry), the T-dual metric is
\begin{equation}
ds^2 = - k (r^2-1) dt^2 + \frac{k dr^2}{(r^2 -
  1)} + \frac{d\theta^2}{k r^2},
\end{equation}
and it is clear that momentum modes will be localised: for example,
the geodesic equation is
\begin{equation}
\dot r^2 - E^2 = \frac{(r^2-1)}{k} (-m^2 -L^2 k r^2). 
\end{equation}
Here we expect that the winding modes of the fundamental string in the
BTZ geometry are localised within an AdS scale of the horizon.

The $B$ field makes it possible for winding modes to propagate to
large $r$ because there is a cancellation between the positive energy
from the tension of the string and a negative contribution to the
energy from the coupling between the string world-sheet and the
background $B$ field. This is the same effect that is responsible for
the existence of long strings in the \ads3 world-sheet theory. If we
have any winding mode which is de-localised on the AdS scale, it has
no potential barrier from moving out all the way to the boundary.

This failure of the tachyon to be localised is a striking result. A
negative consequence is that it will likely be difficult to control
the deformation of the spacetime caused by tachyon condensation.
However, we expect the endpoint of tachyon condensation to be just the
global \ads3 geometry, which would indicate that the tachyon
condensation process only modifies the geometry significantly in the
interior of the spacetime. If this is correct, it may still be
possible to analyse the tachyon condensation. 

\subsection{Milne limit}
\label{milne}

The other flat space limit of interest is near the singularity.
Getting a better understanding of the tachyon in this time-dependent
region is important to understand its effect on singularity
resolution. In this region, the geometry looks locally like a Milne
orbifold of flat space; the generator we are orbifolding along will go
over to a boost generator, rather than a translation generator.
In~\cite{McGreevy:2005ci}, it was argued that there would be a tachyon
localised in the region near the singularity, where the circle is
becoming small. However, this seems to contradict the study of the
Milne orbifold in~\cite{Cornalba:2002fi,Nekrasov:2002kf}, where it was found that
there are no physical states in twisted sectors. On the other hand, it
has been argued that there will be physical states in a different
quantisation of the string~\cite{Pioline:2003bs}. We have physical
twisted sector states in the full BTZ geometry; it is clearly
interesting to ask what happens to them in this limit.

This limit is analogous to the flat-space limit of the elliptic
orbifold in~\cite{Martinec:2001cf}. To make this analogy clear, we
give a brief discussion of that case in \App{ellorb}. The
scalings required to get a regular solution in this limit are
different from in the previous case. We must take $k \to \infty$ with
$r_+$ fixed to get a finite-size identification. The appropriate
coordinates in the limit are $x^2 = \sqrt{k} (t - i \pi/2)$, $\tau =
\sqrt{k} \,r = \sqrt{k} \cosh \rho'$, so we need to take take
$\sqrt{k} \,t$ and $\sqrt{k} \,r$ fixed. Then the metric becomes
\begin{equation}
ds^2 = -d\tau^2 + \tau^2 d\phi^2 + (dx^2)^2 + \CO(1/k),
\end{equation}
where $\phi$ is still a periodic coordinate, $\phi \sim \phi + 2\pi
r_+$.  If we define coordinates $x^3 = \tau \cosh \phi$, $x^1 = \tau
\sinh \phi$, the metric becomes
\begin{equation}
ds^2 = -(dx^3)^2 + (dx^1)^2 + (dx^2)^2,
\end{equation}
and the currents are to leading order simply $J^a = i \sqrt{k}
\partial x^a$, $\bar J^a = i \sqrt{k} \bar \partial x^a$. Thus, the
orbifold is reducing to the usual Milne orbifold in this limit. 

If we took the $B$ field in the gauge \eqref{bfldbtz} and scaled it in
this way, the constant term would blow up. Therefore, we must first
make a gauge transformation to rewrite the $B$ field as
\begin{equation}
B = k r^2 d\phi \wedge dt,
\end{equation}
which becomes 
\begin{equation}
B = \frac{1}{\sqrt{k}} \tau^2 d\phi \wedge dx^2. 
\end{equation}
This vanishes in the limit, but will contribute sub-leading terms to
the equation of motion, as in the previous flat space analysis. 
We again need to keep track of the sub-leading terms in $J^2$, $\bar
J^2$, as we need to consider the terms which are $\CO(1)$ to see the
$\phi$-translation generator. To sub-leading order,
\begin{equation} \label{jmleft}
J^2 = i \sqrt{k} \partial x^2 + i \tau^2 \partial \phi, 
\end{equation}
\begin{equation} \label{jmright}
\bar J^2 = i \sqrt{k} \bar \partial x^2 - i \tau^2 \bar \partial \phi. 
\end{equation}
Again, the Lorentz invariance only implies
\begin{equation}
\bar \partial (\tau^2 \partial \phi) + \partial (\tau^2 \bar \partial \phi)
= 0,
\end{equation}
and we need a sub-leading term in the equations of motion coming from
the $B$ field.  The $x^2$ equation of motion, including this
sub-leading term, is
\begin{equation}
\partial \bar \partial x^2 - \frac{1}{2 \sqrt{k}} (\partial(\tau^2
\bar \partial \phi) - \bar \partial (\tau^2 \partial \phi))  =0 .
\end{equation}
Together with the above equation, this indeed implies the conservation
of $J^2$, $\bar J^2$ to the indicated order.

The important point, however, is that the gauge transformation of the
$B$ field will affect the relation between $J^2_0 - \bar J^2_0$ and
the spacetime energy.\footnote{We thank Eva Silverstein for
  discussions which clarified this point.} In this gauge, we should
define the spacetime energy by \eqref{gens} rather than \eqref{qt2}.
This is clearer from the T-dual perspective. The $B$ field gives rise
to an electric field under dimensional reduction; in the T-dual
\eqref{tdual}, this is the Kaluza-Klein electric field coming from the
metric, and the above gauge transformation is implemented by a
coordinate transformation
\begin{equation}
\theta' = \theta - k t, \quad t' =t. 
\end{equation}
A mode of the scalar field $T$ with energy $\omega$ and momentum $L$
with respect to the original coordinates will have 
\begin{equation} \label{newo}
L' = L, \quad \omega' = \omega + k L
\end{equation}
with respect to these coordinates. Recalling that $L = nr_+$, this is
precisely the difference between \eqref{qt2} and \eqref{gens}, so
$\omega'$ corresponds to the energy \eqref{qt1}. 

Since we hold $\phi$ and $\sqrt{k} t$ fixed as we take $k \to \infty$,
we should take $Q_\phi = J^2_0 + \bar J^2_0 \sim 1$ and $Q_t = J^2_0 -
\bar J^2_0 \sim \sqrt{k}$. The $J^2_0$ ($\bar J^2_0$) eigenvalue for
the twisted sectors is $\lambda + k r_+n/2$ ($\bar \lambda - k r_+
n/2$), so this implies that
\begin{equation} \label{lscale}
\lambda \to \frac{1}{2}(p_\phi + \sqrt{k} p_2 -k r_+ n), \quad \bar
\lambda \to  \frac{1}{2}(p_\phi - \sqrt{k} p_2 +k r_+ n)
\end{equation}
as $k \to \infty$. 

The vertex operators \eqref{twistst} will then have regular limits as
$k \to \infty$. Because the $J^2, \bar J^2$ parts are translation in
$x^2$ (to leading order) in this limit, the boson parts go over to
just a momentum mode vertex operator in the $x^2$ direction.  That is,
from (\ref{jmleft},\ref{jmright}), we see that to leading order, $X
\approx \sqrt{2} \,x^2_L(z)$, $\bar X \approx \sqrt{2} \,x^2_R(\bar z)$,
and \eqref{twistst} becomes
\begin{equation} \label{milnev}
\Phi^n_{j \lambda \bar \lambda} \approx \Psi_{j \lambda} \Psi_{j \bar
  \lambda} e^{-i p_2 x^2}.  
\end{equation}
The parafermion parts represent the dependence on the $x^1, x^3$
directions. For the untwisted sector operators, \eqref{lscale} implies
$\lambda, \bar \lambda \sim \sqrt{k}$, and the parafermions will have finite
dimensions in the limit if $j \sim \sqrt{k}$ as well. This reproduces
the ordinary untwisted sector vertex operators in the limit. Note that
$h(\Psi_{j \lambda}) - \bar h(\Psi_{j \bar \lambda}) = -\(\lambda^2 -
\bar \lambda^2 \)/ k \to 0$ in the limit.

For the twisted sector operators, one might be concerned because the
twist operator~\eqref{twist} is becoming ill-defined in this limit.
This does not prevent us from constructing regular twisted sector
states in the limit. We can regard the twist operator as just a
mathematical device to obtain the physical twisted sector states.
However, this does have an interesting consequence: the twisted sector
states of the orbifold geometry do not arise by twisting the untwisted
sector states surviving the projection. This is because we need different values for $j$ for each sector to get regular parafermion operators in the limit.\footnote{This is similar to the situation arising  in the flat-space limit of the elliptic
orbifold \ads3$/\ZZ_N$, as reviewed in \App{ellorb}.} For the parafermion
parts of the twisted sector operators to remain regular in the Milne
limit, we need to take
\begin{equation}
j \to \frac{1}{2} + \frac{i}{2} ((k-1) r_+ n - \sqrt{k} p_2 + \alpha)
\end{equation}
for some constant $\alpha$,\footnote{The factor of $(k-1)$ multiplying
  $r_+ n$ is introduced for convenience, to cancel a subleading term
  coming from expanding the $(k-2)$ denominator in $h(\Psi_{j
    \lambda}) = -j(j-1)/(k-2) -\lambda^2/k^2$. This would be simply
  $k$ in the superstring case.} so that
\begin{equation} \label{plim1}
h(\Psi_{j \lambda}) \to \frac{1}{2} r_+ n \,\(\alpha + p_\phi\),
\end{equation}
\begin{equation} \label{plim2}
\bar h(\Psi_{j \bar \lambda}) \to \frac{1}{2} r_+ n \,\(\alpha -
p_\phi\). 
\end{equation}
With this scaling, the parafermions should have a regular limit as $k
\to \infty$. These are distinct
from the parafermions arising in the untwisted sector operators. In
particular, we see that
\begin{equation}
h(\Psi_{j\lambda}) - \bar h(\Psi_{j \bar \lambda}) \to r_+ n\, p_\phi.
\end{equation}
This looks like what we would expect for operators carrying $n$ units
of winding and $p_\phi$ units of momentum on a spatial circle, and
indicates that the Milne limit of the BTZ twisted sectors can be
interpreted as describing twisted sectors on the Milne orbifold. This identification is further supported by the fact that the currents $J^\pm$ which reduce to $i \sqrt{k}\,\p x^\pm =i \sqrt{k}\, \p \(x^1 \pm x^3\)$ have the correct monodromies to \req{mono} be twisted sectors of the Milne orbifold. Unlike the flat-space limit of the elliptic orbifolds reviewed in \App{ellorb}, we can choose $j$ so as to get a regular limit for all the twisted sectors. Thus, the spectrum in the Milne limit
includes both the usual untwisted sectors and physical twisted sector
states constructed by the above scaling.

Since we have physical twisted sector states, it would be interesting
to know which of them are tachyonic. Our previous analysis will not be
helpful here, as we restricted our consideration to states with
$\lambda = \bar \lambda$, whereas the twisted sector modes which have a
regular limit have $\bar \lambda - \lambda \sim k r_+ n$. Clearly here
identifying the tachyons will involve disentangling the contribution
to the conformal dimension from winding around the $\phi$ circle.
In this limit as the winding is hidden in the parafermion parts of
the operator, we do not see how to isolate the winding contribution. Perhaps some other representation of the vertex operators will be more helpful here.

For similar reasons, we have difficulty in understanding how localised
these twisted sector modes are. We can attempt to address this
question again from the T-dual point of view. Taking the wave
equation~\eqref{twave} and inserting the change of basis \eqref{newo},
we have
\begin{equation}
  r \partial_r \left( \frac{(r^2- 1)}{k r} \partial_r f \right)
  + \frac{1}{r^2-1} \left( \frac{\omega'^2}{k} - 2 r^2 \omega' L' + k r^2
    L'^2 \right) f = m^2 f.
\end{equation}
Thus, we can see that for modes with $\omega' \sim \sqrt{k}$ and $L'
\sim 1$, near $r=0$ there is a positive contribution to the effective
mass-squared which goes like $k r^2 L'^2$. This should effectively
restrict these modes to the region where $\sqrt{k} r \sim 1$, near the
singularity, as expected by~\cite{McGreevy:2005ci}. 

An important goal for the future is to understand the relation to the
analysis of~\cite{Cornalba:2002fi,Nekrasov:2002kf,Pioline:2003bs}.
In~\cite{Nekrasov:2002kf}, it was argued that a modular-invariant
partition function for the Milne orbifold could be expressed in terms
of a spectrum which only includes untwisted sector states.
In~\cite{Pioline:2003bs}, it was argued that the same partition
function could be given a different interpretation, which involved
scattering states in twisted sectors. Our results are closer to those
of the latter analysis, but this is surprising to us, as the approach
we have adapted on BTZ is a standard quotient construction, and does
not appear to involve any analogue of the non-standard quantisation advocated 
in~\cite{Pioline:2003bs}. Note that we are assuming that parafermionic
operators with the dimensions (\ref{plim1},\ref{plim2}) exist; if no
such regular operators could be constructed, we would be back
with~\cite{Nekrasov:2002kf}. From the BTZ point of view, we would not
expect there to be any problem with the construction of these
parafermion operators, but it should be checked explicitly. These
issues clearly deserve further investigation.

\subsection{Remarks about the spacetime algebra}
\label{asymalg}

It is well-known that asymptotically \ads3 spacetimes have an enlarged
asymptotic symmetry group, which forms two copies of a Virasoro
algebra~\cite{Brown:1986nw}. As a first step towards relating our
perturbative world-sheet study of strings on BTZ to the description in
terms of a dual CFT living on the boundary of the spacetime, it would
be useful to see how this enlarged asymptotic symmetry group emerges
from the world-sheet point of view. For global \ads3, this was
addressed in~\cite{Giveon:1998ns}, where it was shown that the
spacetime \slt $\times$ \slt\ isometries obtained from the world-sheet
currents could be extended to construct the spacetime Virasoro
generators $\CL_n$ by exploiting a special field $\gamma$,\footnote{The field $\gamma$ is the weight zero part of the $\beta-\gamma$ system involved in writing a Wakimoto representation of \slt.} which has
zero conformal dimension and the right charge to fill out the isometry
algebra into a complete Virasoro algebra. This construction is easy to
generalise to elliptic orbifolds of \ads3 as discussed in
\cite{Martinec:2001cf}; for \ads3$/\ZZ_N$ one just keeps the Virasoro
generators $\CL_n$ which are multiples of $N$. These give again a
complete Virasoro algebra. The BTZ spacetime is asymptotically AdS, so
it should be possible to extend the construction to this case as well. This
case is a little more subtle, since we don't have a global \slcur{k}
to provide clues; the orbifold action leaves only a $\widehat{U(1)}$
algebra. Also, the algebra will not arise as a restriction of the
Virasoro algebra of the covering space in this case, as none of those
generators commute with the orbifold action. As a result, all that we
can do is to suggest the form that the Virasoro generators should
take.

We assume that the construction will proceed in much the same way as
in the \ads3 case \cite{Giveon:1998ns}, identifying a physical vertex
operator that has dimension zero and $J^2$ charge $1$, to play the
role of the field $\gamma$. The monodromies of the currents in
the $n^{{\rm th}}$ twisted sector are
\begin{equation}
J^2(e^{2\pi i} z) = J^2(z), J^\pm(e^{2\pi i} z) = e^{\mp
    2\pi r_+ n} J^\pm(z),
\Label{mono}
\end{equation}	
which could be realised by giving the free boson $X$ \req{hyppara1}
monodromy $X(e^{2\pi i} z) = X(z) - 2\pi r_+ n\, \sqrt{\frac{k}{2}} $. This
would imply that the monodromies of the untwisted sector vertex
operators are
\begin{equation}
\Phi_{j\lambda}(e^{2\pi i} z) = e^{2\pi i r_+ n \lambda} \Phi_{j
    \lambda} (z).
\Label{mono2}
\end{equation}	
The spacetime $\CL_0$ generator is $\CL_0 = - r_+ \oint dz J^2(z)$, where
we have introduced a normalisation factor $r_+$, which is required to
make the charges work out correctly, but perhaps also seems natural
from the spacetime point of view. With this normalisation, the
spacetime symmetry generators $\CL_0-\bar \CL_0$ will generate angular
momentum with respect to the $2\pi$ periodic coordinate
$\phi/r_+$.\footnote{The normalisation was fixed in the AdS case by
  considering the global \slt.}  We want the $\CL_n$ to have $\CL_0$
charge $n$, and we need to integrate a well-defined (trivial
monodromy) world-sheet operator of conformal dimension one.  We see
that $\Phi_{0 (m/r_+)}$ has the right properties to be identified as
$(\gamma_{BTZ})^m$, so an appropriate ansatz is
\begin{equation}
\CL_n = - r_+ \oint dz \left[ f_1(n) J^2 \Phi_{0 (n/r_+)} + f_2(n) J^- \Phi_{0
    (n/r_+ +i)} + f_3(n) J^+ \Phi_{0 (n/r_+ -i)} \right], n \in \mathbb{Z}
\Label{virasy}
\end{equation}	
Note the factors of $\pm i$ in the vertex operators, which are
required to cancel the $J^2$ charges of $J^\pm$. The functions
$f_i(n)$ are to be fixed by the requirement that the algebra of the
$\CL_n$s closes correctly into a Virasoro algebra.

Morally, this is how the spacetime Virasoro algebra should arise from
the world-sheet point of view. To check this in detail, we would need
to know the OPEs of the primary operators $\Phi_{0 \lambda}$ to
evaluate the commutators and work out the appropriate choices for the
coefficients, which we leave as an interesting exercise for the
future. We postpone further discussion of the relation of our
world-sheet analysis to the dual CFT point of view to the discussion
in \sec{discuss}.

\section{The superstring}
\label{sstring}

We would now like to extend our discussion to the superstring. This
will eliminate the bulk tachyon of the bosonic theory; as noted
before, our winding tachyons are not well localised, so it is still
difficult to obtain control of the decay of our spacetime, even after
eliminating the bulk tachyon. However, we consider it useful to verify
that there is a GSO projection which eliminates the tachyon in the
untwisted sector but retains it in twisted sectors. For BTZ, there are
two possible choices of spin structure, and we will see that the
tachyon in odd twisted sectors survives the GSO projection if we take
an anti-periodic spin structure on spacetime. Also, the world-sheet
theory has some interesting technical features. We will consider Type
II string theory on BTZ $\times \Sp^3 \times {\bf T}^4$ for
simplicity.

\subsection{The superstring WZW model}
\label{susyworlds}

The superstring on \ads3 is described by a \slcur{k} super-WZW
model~\cite{Giveon:1998ns}. We begin by reviewing some aspects of this
model in the hyperbolic basis, which is adapted to the orbifold we
want to consider. The world-sheet WZW model with \slcur{k} current
algebra has generators\footnote{In this section the total current
  including the fermionic contribution will be denoted as $J^a$; since
  we will no longer talk about the bosonic theory this notation should
  hopefully cause no confusion.} $J^a$ at level $k$. This can be
decomposed into a bosonic \slcur{k+2} at level $\tk = k+2$, whose
generators we denote as $j^a$, and a set of free fermions. Our
conventions for the super-current algebra are
\begin{equation}
J^a = j^a - {i \over k} \, \eps^a_{\ b c} \, \psi^b\, \psi^c  ,
\Label{cursusy}
\end{equation}	
with
\begin{eqnarray}
\psi^a (z) \, \psi^b(w) & \sim & \frac{k}{2} {\eta^{ab} \over (z-w)}, \nonu
j^a(z) \,\psi^b(w) & \sim & 0,
\Label{opesusy}
\end{eqnarray}	
and as before
\begin{equation}
j^a(z) \,j^b(w) \sim {\tk\over 2} \, {\eta^{ab} \over (z-w)^2 } + i \,
{\eps^{a b}_{\ \ c} \, j^c \over (z-w)}, 
\end{equation}
so
\begin{equation}
J^a(z) \,J^b(w) \sim {k\over 2} \, {\eta^{ab} \over (z-w)^2 } + i \,
{\eps^{a b}_{\ \ c} \, J^c \over (z-w)}. 
\end{equation}
The world-sheet $\CN =1$ super-current in these conventions is then given as
\begin{equation}
G(z) = {2\over k} \, \(g_{ab}\, \psi^a \, j^b - {i\over 3 k} \, \eps_{abc} \, \psi^a\, \psi^b\, \psi^c \) . 
\Label{supcur}
\end{equation}	
Our conventions for the \slc\ are $\eps^{123} =1$ and $\eta_{ab}=  {\rm diag} (1,1,-1)$

The internal CFT which has target space $\Sp^3 \times {\bf T}^4$ will
have more of a role in the superstring than previously, as we need to
work out the appropriate spin fields. The $\Sp^3$ part is a
world-sheet $\CN=1$ $SU(2)_k$ WZW model while the ${\bf T}^4$ is a free $\CN =1$ SCFT. The $\widehat{SU(2)}_k$ algebra is generated by
(again $K^a$ are the total currents and the $k^a$ represent the
bosonic contribution)
\begin{equation}
K^a = k^a - {i \over k} \, \eps^a_{\ b c} \, \chi^b\, \chi^c ,
\Label{cursususy}
\end{equation}	
with 
\begin{eqnarray}
\chi^a (z) \, \chi^b(w) & \sim& \frac{k}{2} {g^{ab} \over (z-w)}, \nonu
k^a(z) \chi^b(w) &\sim& 0, \nonu
k^a(z) \,k^b(w) &\sim &{\tk\over 2} \, {g^{ab} \over (z-w)^2 } + i \,
{\eps^{a b}_{\ \ c} \, k^c \over (z-w)},
\Label{opesusy2}
\end{eqnarray}	
where now $\tk = k-2$, so
\begin{equation}
K^a(z) \,K^b(w) \sim {k\over 2} \, {g^{ab} \over (z-w)^2 } + i \,
{\eps^{a b}_{\ \ c} \, K^c \over (z-w)},
\end{equation}
and  the world-sheet super-current is 
\begin{equation}
G(z) = {2\over k} \, \(g_{ab}\, \chi^a \, k^b - {i\over 3 k} \, \eps_{abc} \, \chi^a\, \chi^b\, \chi^c \).
\Label{supcur2}
\end{equation}	
Of course, the major difference is that the metric is positive definite: $g_{ab} = {\rm diag} (1,1,1)$. 

\paragraph{Bosonisation of the free fermions:} To write down spin
fields it is useful to bosonise the fermions. Since we wish to work in
the hyperbolic basis we want to diagonalise $J^2$ for the \slc\
current algebra. From the definition of the total current
\req{cursusy}, we see that the fermionic current involved in $J^2$ is
made up of $\psi^1\, \psi^3$ so we want to bosonise this combination
into a single free boson. The natural extension of the story for the
elliptic basis of \cite{Giveon:1998ns} would be to consider the
following bosonisation rules:
\begin{eqnarray}
\; \, \p H_1 & =& J^2 - j^2 = - { 2\,i \over k } \, \psi^1 \, \psi^3, \nonu 
i \, \p H_2 & = &K^2 - k^2 = + { 2\, i \over k } \, \chi^1 \, \chi^3, \nonu
i\, \p H_3 & = &- { 2\, i \over k} \, \psi^2 \, \chi^2, \nonu
i\, \p H_4 & =& - i \, \lam^1 \, \lam^2, \nonu
i\, \p H_5 &= &-i\, \lam^3 \, \lam^4,
\Label{bosrules}
\end{eqnarray}	
where the $\lam^i$ are the free fermions for the ${\bf T}^4$ part of
the story, and the bosons $H_i(z)$ are all canonically normalised
\begin{equation}
H_i(z) \, H_j(w) = - \d_{ij} \, \log (z-w).
\Label{Hope}
\end{equation}	
For future reference we also give the expression for the fermions
directly in terms of the bosonic fields, 
\begin{eqnarray}
\psi^3 & = &-i \, {\sqrt{k} \over 2} \, \(e^{i\, H_1} + e^{-i\, H_1} \), \nonu
\psi^1 & = & i \,  {\sqrt{k} \over 2} \, \(e^{i\, H_1} - e^{-i\, H_1} \), 
\Label{fermirepH}
\end{eqnarray}	
which imply that 
\begin{equation}
\psi^\pm \equiv \psi^1 \pm \psi^3 = \mp i\,\sqrt{k} \, e^{\mp i\, H_1}.
\Label{fermirecomb}
\end{equation}	

\paragraph{The super-parafermions:} As in the discussion of the
bosonic string we find it useful to work with a super-parafermion
representation of the \slcur{k} and the $\widehat{SU(2)}_k$ current
algebras. Concentrating on the \slcur{k} current algebra we introduce
a bosonic representation for the currents,
\begin{equation}
J^2(z) = - i \, \sqrt{{k\over 2}} \, \p \CX.
\end{equation}
As before, we also have a bosonic representation for the bosonic
current, 
\begin{equation}
j^2(z) = -i\, \sqrt{{\tk \over 2}} \, \p X, 
\Label{boscur}
\end{equation}	
and the bosons $\CX$ and $X$ are both canonically normalised, so $\CX(z) \, \CX(w) = -\log (z-w)$. Clearly, by virtue of \req{cursusy} we have 
\begin{equation}
i \, H_1 = \sqrt{{k\over 2}} \, \CX - \sqrt{{\tk \over 2}} \, X.
\Label{bosrelA}
\end{equation}	
It is useful to introduce another canonically normalised boson,
$\CH_1(z)$, which is orthogonal to $\CX$, so that we can write
\begin{eqnarray}
H_1 &= i \, \sqrt{{2 \over k}} \, \CX + \sqrt{{\tk \over k}} \, \CH_1, \nonu 
X & = \sqrt{{\tk \over k}} \, \CX - i \, \sqrt{{2 \over k}} \, \CH_1.
\Label{boscurrel}
\end{eqnarray}	
Note that in the flat space limit $k \to \infty$, $\CX = X$ and $\CH_1
= H_1$.

The remainder of the currents $j^\pm$ are written by introducing
parafermions
\begin{equation}
j^\pm = j^1 \pm j^3 = \xi^\pm \, e^{\pm \sqrt{{2\over \tk}} \, X}
= \xi^\pm \,  e^{\pm \sqrt{{2\over k}} \, \( \CX - i \, \sqrt{{2 \over \tk}} \, \CH_1\)} ,
\Label{bospara}
\end{equation}		
using the fact that $j^\pm$ carry imaginary $j^2$ charge $\pm \, i$
respectively.  The fermions which are bosonised as in \req{fermirepH}
can be written in terms of the bosons $\CX, \CH_1$ as:
\begin{equation}
\psi^\pm = \psi^1 \pm \psi^3 = \mp\,i\, \sqrt{k} \, e^{\mp \, i \, H_1} = \mp \, i\,\sqrt{k}\, e^{\mp \(i\,\sqrt{\, {\tk \over k}} \, \CH_1  - \sqrt{{2 \over k}} \, \CX\) }.
\Label{fermirepT}
\end{equation}	
Note that this implies that the fermions carry imaginary $J^2$ charge.
In the hyperbolic basic one linear combination of the $J^2$ charge
measures the spacetime energy. Fermions in this basis therefore have
imaginary spacetime energy! This is a consequence of the
transformation properties of the spacetime fermions and vector fields
under the hyperbolic generator of \slt.

Finally, we can write down the supercurrent in terms of the
parafermion representation used above,
\begin{equation}
\sqrt{k} \, G(z) =   i\, \xi^+ \, e^{i\,\sqrt{{k\over \tk}} \, \CH_1}  -i\,  \xi^- \, e^{-i\,\sqrt{{k\over \tk}} \, \CH_1}  -  \sqrt{2} \, i \,\psi^2 \, \p \CX.
\Label{supercbos}
\end{equation}	
As in the elliptic basis, the boson $\CX$ associated with the total
current only appears differentiated in the expression for the
supercurrent. This implies that the supercurrent will be mutually
local with respect to the twist operator we will introduce to
implement the orbifold.

\paragraph{The spin fields:} The simplest set of spin fields we can
write down are
\begin{equation}
S_\a = e^{{i\over 2} \, \eps_I \, H_I} \ , 
\Label{spinflds}
\end{equation}	
where the $H_I$ are the canonically normalised bosons introduced in
\req{bosrules}. Note that the spin fields only involve $H_1$, and not
$\CH_1$. To determine the OPE with the world-sheet supercurrent $G(z)$
given in \req{supercbos} is straightforward. The most singular terms
in the OPE come from the three fermion piece -- this has to cancel to
ensure that the $G(z) \, S_\a(w)$ OPE has as its leading singularity a
square root branch cut. This calculation works along the same lines as
in \cite{Giveon:1998ns}, with the $(z-w)^{{3\over 2}}$ singularity
being cancelled by an interplay between the contributions from the
\slcur{k} part and the $\widehat{SU(2)}_k$ part. This leads to the
condition derived by \cite{Giveon:1998ns},
\begin{equation}
\prod_{I =1}^3\, \eps_I = 1 .
\Label{spinG}
\end{equation}	
Furthermore the $S_\a(z)\, S_\b(w)$ OPE is local provided 

\begin{equation}
\prod_{I=1}^5 \, \eps_I = 1.
\Label{spinspinOPE}
\end{equation}	
To do this calculation it is useful to write down a para-fermionic
representation for the $\widehat{SU(2)}_k$ theory as well; up to some
signs and factors of $i$ one defines $\(\CY, Y, \CH_2, H_2\)$ which are
analogous to set of bosons $\(\CX, X, \CH_1, H_1\)$ described above.
For details see~\cite{Martinec:2001cf}.  The conditions
(\ref{spinG},\ref{spinspinOPE}) define the set of spin fields on \ads3
$\times \Sp^3 \times {\bf T}^4$. Note that while our focus is on a
particular choice of internal CFT, the considerations here can be
easily generalised to a more general internal space as
in~\cite{Giveon:1999jg}.

\paragraph{Vertex operators:} The NS ground states are 
\begin{equation}
{\mathcal T}_{j j'} = e^{-\varphi} \Phi^{SL(2)}_{j\lambda}
\Phi^{SU(2)}_{j' m} e^{i q\cdot Y}, 
\end{equation}
where $\Phi^{SL(2)}_{j\lambda}$ is a primary operator of the \slcur{}
current algebra considered in the bosonic string discussion \req{chiralpara},
$\Phi^{SU(2)}_{j' m}$ is a primary of the $\widehat{SU(2)}$ algebra
associated with the $\Sp^3$, $e^{i q\cdot Y}$ represents the winding
and momentum on the $T^4$, and $\varphi$ is the bosonised
super-reparametrisation ghost. These operators have dimension
\begin{equation}
h = \frac{-j(j-1)}{k} + \frac{j' (j'+1)}{k} + \frac{q \cdot q}{2}. 
\end{equation}
The first excited states in the NS sector are constructed by adding a
world-sheet fermion to this operator, so for example
\begin{equation}
{\mathcal V}^i_{j j'} = e^{-\varphi} \lambda^i \Phi^{SL(2)}_{j\lambda}
\Phi^{SU(2)}_{j' m} e^{i q\cdot Y}.
\end{equation}
The R ground states are  constructed by adding a spin field,
\begin{equation}
{\mathcal Y}_R = e^{-\varphi/2} S \Phi^{SL(2)}_{j\lambda}
\Phi^{SU(2)}_{j' m} e^{i q\cdot Y}.
\end{equation}
A detailed discussion of which states survive in the BRST cohomology,
and of the charges associated with the R states, can be found
in~\cite{Argurio:2000tb}. Since we are interested in the tachyons, we
will be content with observing that the only physical vertex operators
involving continuous representations of \slt\ are NS ground states. 

\subsection{The BTZ orbifold}
\label{susybtz}

We can now consider the orbifold of the \slcur{k} super-WZW model to
construct the BTZ spacetime in the superstring. We begin by presenting
a twist field that implements the BTZ orbifold projection and then go
on to discuss the GSO projection involved in the superstring, showing
that for anti-periodic spin structure, the projection retains the
tachyon in odd twisted sectors. 

In the superstring, it is natural to construct the twist field for the
BTZ orbifold from the total \slcur{k} current $\CX$, as in the
elliptic case~\cite{Martinec:2001cf}. We therefore consider
\begin{equation}
t_n = e^{i \, r_+ \, n \, \sqrt{{k\over 2}}\, \(\CX -\bar{\CX}\) }.
\Label{twistsy}
\end{equation}	
As mentioned previously, the super-current $G(z)$ \req{supercbos} is
mutually local with respect to this twist operator, so the orbifold
will preserve the world-sheet supersymmetry.\footnote{One might also
  argue as in \cite{Martinec:2001cf} that the boundary Virasoro
  algebra (\ie, the spacetime theory currents) is generated from the
  total currents $J^a$ and not the bosonic currents $j^a$. For the
  hyperbolic \slcur{k} unlike the elliptic case we don't have a clean
  expression for the spacetime algebra, but one expects the structure
  to be maintained. See later for a discussion on this issue.}  This
performs the same twist as before on the bosonic currents. The
additional part of the twist involving the field $H_1$ which
bosonises the spinor fields can be understood as implementing the
correct transformation properties for the spacetime spinor and vector
indices (which are carried by the world-sheet fermions) under the \slt\
generator that we are orbifolding by.

The spin fields will not be mutually local with the twist operator
\eqref{twistsy}: the $t_n(z) \, S_\a(w)$ OPE has a logarithmic branch
cut. Thus, spacetime supersymmetry is completely broken by the
orbifolding, as we would expect.  We can construct spacetime fermions
by combining $S_\a(w)$ with a bosonic vertex operator with an
imaginary value for $\lambda + \bar \lambda$, producing a compensating
branch cut to give a R-NS vertex operator which is mutually local wrt
$t_n$. These correspond to the modes of the spacetime fermions which
are invariant under the combined action of the translation and an \slc\
rotation of the spinor indices.

We can now implement the orbifold and obtain the twisted sector
states. Naively, we should proceed as in the bosonic case, imposing
mutual locality with respect to the twist operator \eqref{twistsy} and
including all the twisted sector states required to achieve closure of
the OPE. However, there is a small subtlety: in the superstring, the
states in untwisted sectors are not mutually local until we impose the
GSO projection. They can have square root branch cuts in the OPE.
Therefore, at this stage we may need to allow square root branch cuts
in the OPE with $t_n$. Having constructed a general set of twisted
sectors in this way, we will seek a GSO projection which gives a
mutually local spectrum.

For the NS-NS states, the OPE with the twist operator \eqref{twistsy}
will give
\begin{equation}
t_n \mathcal T \bar{\mathcal T} \sim \frac{1}{(z-w)^{r_+ n (\lambda + \bar \lambda)}}, 
\end{equation}
so demanding mutual locality of these operators with respect to the
twist operator imposes $r_+ (\lambda + \bar \lambda) \in \mathbb Z$,
as in the bosonic case, quantising the momentum around the circle. For
the R-NS states, the OPE with the twist operator gives
\begin{equation}
  t_n \mathcal Y_{R} \bar{\mathcal T} \sim \frac{1}{(z-w)^{r_+ n
      [(\lambda + \bar \lambda) - \frac{i}{2}]}}. 
\end{equation}
As explained earlier, the factor of $i$ here comes from the
transformation properties of spacetime spinors under the hyperbolic
generator which we orbifold along. Requiring mutual locality with
respect to $t_n$ will then impose $r_+ [(\lambda + \bar \lambda) -
\frac{i}{2}] \in \mathbb{Z}$, which corresponds to choosing modes of
the spacetime spinor which are invariant under the orbifold action.
However, in the BTZ spacetime, there are two possible choices of spin
structure; since the $\partial_\phi$ circle is not contractible,
fermions can be either periodic or anti-periodic around this circle.
Considering anti-periodic fermions corresponds to imposing $r_+
[(\lambda + \bar \lambda) - \frac{i}{2}] - \half \in \mathbb{Z}$,
giving a square root branch cut in the OPE with $t_n$. For the R-R
states, the OPE with the twist operator gives
\begin{equation}
  t_n \mathcal Y_{R} \bar{\mathcal Y}_{R} \sim \frac{1}{(z-w)^{r_+ n
      [(\lambda + \bar \lambda) - i]}}, 
\end{equation}
and we should impose mutual locality to obtain $r_+ [(\lambda + \bar
\lambda) - i] \in \mathbb{Z}$, corresponding to choosing modes of
the spacetime fields which are invariant under the orbifold
action. The analysis for the excited NS-NS states is similar. 

Twisted sector states are constructed by taking the composite
operators arising from the product of the $t_n$ with the invariant
untwisted sector operators. For the NS-NS ground states, the twisted
sector operators are
\begin{equation} \Label{nstwist}
{\mathcal T}^n_{j,j'} = e^{-\varphi} e^{-\bar \varphi}
\Phi^{SL(2)}_{j\lambda} \bar \Phi^{SL(2)}_{j \bar \lambda}
\Phi^{SU(2)}_{j' m} \bar \Phi^{SU(2)}_{j' \bar m} e^{i q\cdot Y} e^{i
  \bar q \cdot \bar Y} e^{i \, r_+ \, n \, \sqrt{{k\over 2}}\, \(\CX
  -\bar{\CX}\) }. 
\end{equation}
To calculate the dimensions of these operators, it is useful to
rewrite them in terms of parafermions or super-parafermions, but we
will not do so explicitly here; the construction closely parallels the
elliptic case discussed in~\cite{Martinec:2001cf}. The dimensions of
these operators are
\begin{equation}
h = \frac{-j(j-1)}{k} + \frac{j' (j'+1)}{k} + \frac{q \cdot q}{2} -
\frac{\lambda^2}{k} + \frac{(\lambda + k r_+ n/2)^2}{k}, 
\end{equation}
\begin{equation}
\bar h = \frac{-j(j-1)}{k} + \frac{j' (j'+1)}{k} + \frac{q \cdot q}{2} -
\frac{\bar \lambda^2}{k} + \frac{(\bar \lambda - k r_+ n/2)^2}{k}. 
\end{equation}
We adopt the same definition of a tachyon as in the bosonic case, so a
mode with $\lambda = \bar \lambda$ is considered tachyonic if and only
if $-j(j-1) - \lambda^2 >\frac{1}{4}$, so that the \slcur{}
superparafermion part of the operator is of sufficiently positive
dimension. As in the untwisted sector, only the NS-NS ground states
can be both physical states and tachyonic; in the other sectors, the
positive contribution to the conformal dimension from the fermions or
spin fields makes it impossible to satisfy the physical state
condition for $-j(j-1) - \lambda^2 > \frac{1}{4}$. For large $k$, we
can find tachyons which satisfy the physical state condition $h= \bar
h = \frac{1}{2}$ only if $\sqrt{k} r_+ < \sqrt{2}$.

Turning to the GSO projection, we assume that we make the standard
projection in the untwisted sector, projecting out the ground states
in the NS-NS sector, and defining a mutually local set of operators in
the untwisted sector. In the case where the R-NS vertex operators are
mutually local with respect to $t_n$, corresponding to the periodic
spin structure on spacetime, this extends trivially to the twisted
sectors, projecting out the NS-NS ground states in every sector. By
contrast, when the R-NS vertex operators have a square root branch cut
with respect to $t_n$, corresponding to the anti-periodic spin
structure on spacetime, the NS-NS ground states in odd twisted sectors
are mutually local with respect to the states we keep in the untwisted
sector, so they will be retained under GSO projection.  In summary,
when we choose an anti-periodic spin structure for the fermions on
spacetime, the tachyons which survive the GSO projection are
\eqref{nstwist} for the odd twisted sectors.

In the flat space limit of \sec{btzflatnh}, this GSO
projection reduces to the usual Scherk-Schwarz GSO projection on the
translational orbifold, so we recover the usual flat space analysis in
this limit. In the Milne limit of \sec{milne}, the twist
operator \eqref{twistsy} becomes ill-defined, as previously noted.
However, since the vertex operators have regular limits, we should be
able to consider either choice of GSO projection in this limit. Thus,
there should exist a GSO projection on the Milne orbifold
corresponding to an antiperiodic spin structure on the orbifold, in
which we keep the NS-NS ground states in odd twisted sectors.

Finally, let us remark on the asymptotic symmetry algebra for the
superstring. As in the bosonic theory, the asymptotic symmetries of
asymptotically \ads3 spaces are enlarged to two copies of the Virasoro
algebra. From the dual CFT point of view, we would expect this to now
be embedded in a superconformal algebra. In~\cite{Argurio:2000tb}, the
extension of the spacetime supersymmetry to obtain the full set of
asymptotic super-isometries from the world-sheet point of view was
sketched. In the present case, the spacetime supersymmetry is broken,
but we would expect the BTZ geometry will have the same asymptotic
superconformal symmetry algebra, since the spacetime is still
asymptotically \ads3, and hence has asymptotic Killing spinors. It
should be possible to construct this asymptotic super-isometry algebra
from the world-sheet point of view following~\cite{Argurio:2000tb} and
our discussion of the Virasoro algebra in \sec{asymalg}, but
we will not explore this further here. 
It would be interesting to understand the relation our construction to the asymptotic super-isometry algebra constructed from a  supergravity  point of view (along the lines of~\cite{Brown:1986nw}) by~\cite{Henneaux:1999ib}.

\section{Discussion}
\label{discuss}

We have studied the closed string tachyons on the BTZ black hole with
NS-NS flux, by treating it as an orbifold of \ads3. We used a
parafermion representation of the current algebra. We found that for
the superstring, there is no closed string tachyon if we choose the
spacetime spin structure which imposes periodic boundary conditions
for fermions on the spatial circle. For the spin structure with
anti-periodic boundary conditions, we showed that there is a tachyon
in odd twisted sectors if the proper size of the circle at the event
horizon is small enough. We focused on operators with $\lambda=\bar
\lambda$, which corresponds to zero spacetime energy in the usual
gauge, and argued that the appropriate definition of a tachyon for
such operators was that the conformal dimension of the parafermionic
part of the operator should be positive. In the superstring, this
condition can be satisfied if $\sqrt{k} r_+ < \sqrt{2} \l_s$.

Surprisingly, this tachyon is not localised in the region where the
spatial circle is small. The wavefunctions for twisted sector states
have the same radial falloff as for the corresponding untwisted sector
states. This is due to coupling to the background $B$ field, which
cancels the positive energy coming from stretching the string as the
circle becomes large. That is, the tachyon is a long string mode, and
as such, can propagate out to infinity. It would be interesting to
examine the asymptotically flat black strings constructed from the
F1-NS5 system, which approach this geometry in the near-horizon limit:
our results suggest that the tachyonic winding strings in these
backgrounds should be localised near the black string, but on a scale
set by the charges, rather than in the much smaller region where the
circle they wrap is of order the string scale.

Since the failure of this tachyon to be localised is associated with
the presence of NS-NS flux, one might hope to construct an example in
which it is localised by considering instead a BTZ black hole with R-R
flux. We should note first that this is considerably more difficult
than the present case, as the WZW methods we have used here will not
be available.  It is possible that one can use the D1-D5 world-sheet
CFT~\cite{Berkovits:1999im} and show that the BTZ geometry in that
system indeed has a tachyon when the horizon size is less than the
string scale, and that the tachyon wavefunction is supported within an
AdS radius. It may also be possible to make some progress by studying
the supergravity spectrum in the T-dual geometry. However, one can
also observe that the BTZ black hole with R-R flux is S-dual to the
case we have considered here, so even if the winding tachyons of the
fundamental string on that background are localised near the horizon,
one would expect it to have instabilities to the condensation of
winding D-strings, S-dual to the fundamental strings we have
considered, and by our analysis, this D-string instability will not be
localised in the near-horizon region.

One could also consider simpler single-charge black string solutions,
where the near-horizon region is not a BTZ black hole. Here we have no
reason to expect that the winding tachyon will not be localised in the
near-horizon region, as suggested by an approximate analysis. One
might take our results as counselling caution in over-reliance on such
approximate arguments, however. A more careful analysis in such cases
will be quite difficult.

There are a number of directions for further investigation arising
from this work. Although our tachyon is not localised, it is clearly
important to try to understand its condensation. The natural endpoint
for condensation of the twisted sector tachyons on BTZ is the global
\ads3 spacetime. Since this only requires a change in the geometry in
the interior of the spacetime, there is some hope that we can gain
some insight into the tachyon condensation process. Perhaps tachyon
condensation produces $\CO(1)$ changes in the geometry everywhere,
which are negligible compared to the $\CO(r^2)$ behaviour of the
background metric at large distances.

It is also important to understand in detail the relation between the
Milne limit of our spectrum and the spectra calculated directly in
flat space in~\cite{Nekrasov:2002kf,Pioline:2003bs}. In particular, it
would be useful to have a more explicit description of the Milne limit
of the twisted sector vertex operators. It may be that adopting a
different realisation of the current algebra, such as the Wakimoto
representation used in~\cite{Satoh:1997xe,Giveon:1998ns}, would be helpful.

It would also be interesting to extend our analysis to other orbifolds
of \ads3. First, we could extend our work to the rotating BTZ black
hole, which would involve considering an asymmetric orbifold, which
acts differently on left- and right-movers. In this case, it would be
difficult to rigorously establish modular invariance, but one can
simply extend our analysis by introducing an appropriate twist
operator, and hope that the resulting spectrum is modular invariant.
The main example of interest is the supersymmetric BTZ black hole,
which correspond to orbifolds of \ads3 by a parabolic generator. If we
choose the supersymmetry-breaking spin structure on this spacetime, we
expect to have a winding tachyon. This case is analogous to the
quotient of AdS$_5$ considered in~\cite{Horowitz:2006mr}, and near the
singularity, it will approach the null orbifold of flat space, so
there are interesting connections to explore here. Another interesting
extension would be to consider the ``swedish geons''
\cite{Aminneborg:1997pz}, which are BTZ-like orbifolds of \ads3 with a
single exterior region. These can potentially provide examples where
the tachyon condensation can lead to disconnected target space
geometry with intricate topology, essentially providing a rich set of
examples to probe baby universes in string theory.

The other central issue for future development is to understand the
description of this tachyon and its condensation in the dual CFT on
the boundary of the spacetime. We have tried to make some first steps
in this direction by exploring the construction of the asymptotic
isometry algebra from the world-sheet point of view. However, there are
significant barriers to going further: First, we do not understand the
theory on the F1-NS5 worldvolume, so there is no first principles
construction of the dual CFT. Second, we do not know how to interpret
the twisted sectors, which correspond to long strings wrapping the
spatial circle in the boundary, from the dual CFT point of view. This
is a problem even in pure \ads3. From a technical point of view, in
\cite{Giveon:1998ns} it was shown that the spectral flowed vertex
operators have unconventional transformation properties with respect
to the Virasoro algebra of the dual CFT.  It will be an interesting
problem for the future to make progress on the interpretation of the
construction of the spacetime Virasoro algebra and the behaviour of
physical twisted sector vertex operators from this perspective.

\section*{Acknowledgements}
\label{acks}
It is a pleasure to thank Micha Berkooz, Jan de Boer, Gary Horowitz,
Veronika Hubeny, Nori Iizuka, Esko Keski-Vakkuri, James Lucietti,
Shiraz Minwalla, Moshe Rozali and Eva Silverstein for useful
discussions. This work was supported in part by EPSRC and PPARC, and
by the NSF through its support for the Aspen Centre for Physics.  We
would like to thank Aspen Centre for Physics for hospitality during
the course of this project. MR would also like to acknowledge the
hospitality of KITP, Santa Barbara, during the Quantum Nature of Spacetime Singularities mini-program.

\appendix

\section{The \slt\ Lie algebra}
\label{ads3btz}
The group \slt\ is the group of $2 \times 2$ matrices with unit
determinant and the elements taking real values. The Lie algebra of
\slt\ is given by the commutation relations:
\begin{equation}
\comm{T^a}{T^b} = \epsilon^{ab}_{\ \ c} \, T^c \  ,
\Label{liesl}
\end{equation}	
where $\epsilon^{123} = 1$, and the index is lowered with the metric
$\eta_{ab} = \mbox{diag} (1,1,-1)$. Explicitly, one can choose a
representation in terms of Pauli matrices:
\begin{equation}
T^1 = {1\over 2} \, \sigma^3 \ , \; \;\; \; T^2 =
  {1\over 2} \sigma^1 \;\;\;\; T^3 = - {i\over 2} \, \sigma^2 \ . 
\Label{paulidef}
\end{equation}	
We will however find it convenient to work with a different set of
generators $\tau^a = i \, T^a$ in terms of which we can express
\req{liesl} in a more familiar form
\begin{equation}
 \comm{\tau^a}{\tau^b} = i\, \epsilon^{ab}_{\ \ c} \,\tau^c ,
\Label{tauliesl}
\end{equation}	
similar to the $SU(2)$ commutation relations. This version is more
appropriate because when \slt\ $\times$ \slt\ occurs as the isometry
algebra of \ads3, we are interested in real eigenvalues for the
generators $\tau^a$. For the elliptic basis, which is natural when
thinking of the group as $SU(1,1)$, we take $\etau^3 = \tau^3$ and
$\etau^\pm = \tau^1 \pm i \tau^2$, so one has the commutation
relations
\begin{equation} [\etau^3, \etau^\pm] = \pm \etau^\pm\ , \;\;
  [\etau^+, \etau^-] = -2 \, \etau^3.
\Label{ellalg}
\end{equation}	
On the other hand, if we think of the \slt\ description, it is more
natural to use the hyperbolic basis, $T^\pm = T^1 \pm T^3$, in terms
of which the commutation relations are
\begin{equation}
[T^2, T^\pm] = \pm  T^\pm\ , \;\;  [T^+ , T^-] = 2\, T^2 \ .
\Label{hypalg}
\end{equation}	
Note that $T^\pm$ are not related to $\etau^\pm$, despite the
similarity in the commutators. Again, even when working in the
hyperbolic basis, we use the $\tau^a$, not $T^a$. The quadratic
Casimir of \slt\ is 
\begin{equation}
c_2 =- (\tau^3)^2 + (\tau^1)^2+ (\tau^2)^2 .
\Label{quadcasi}
\end{equation}	
%

\section{Flat-space limit of the elliptic orbifold}
\label{ellorb}

In studying the Milne limit of our orbifold, we found it useful to
compare to the flat-space limit of the elliptic orbifold studied
in~\cite{Martinec:2001cf}. Since the flat-space limit is not discussed
very explicitly in that reference, we give some formulae here to
enable comparison with \sec{milne}. The elliptic orbifold is
the quotient of global \ads3 \eqref{globalmet} by the $\mathbb{Z}_N$
group generated by $\theta \to \theta +
2\pi/N$. In~\cite{Martinec:2001cf}, this was studied by bosonising
$J^3$, 
\begin{equation}
J^3 = - \sqrt{\frac{k}{2}} \partial X, \quad J^\pm = \xi^\pm e^{\pm
  \sqrt{\frac{2}{k}} X},
\end{equation}
and adopting a parafermionic representation for the current algebra
primaries,
\begin{equation}
\Phi_{j m \bar m} = \Psi_{j m \bar m} e^{\sqrt{\frac{2}{k}} (m X +
  \bar m \bar X)}. 
\end{equation}
The dimensions of the parafermions $\Psi_{j m \bar m}$ are
\begin{equation}
h = -\frac{j(j-1)}{(k-2)} + \frac{m^2}{k}, \quad \bar h = -
\frac{j(j-1)}{(k-2)} + \frac{\bar m^2}{k}. 
\end{equation}
The orbifold can then be implemented by introducing a twist operator
\begin{equation}
t_w = e^{\frac{q}{N} \sqrt{\frac{k}{2}} (X + \bar X)}.
\end{equation}
This gives twisted sector operators
\begin{equation}
\Phi^q_{j m \bar m} = \Psi_{j m \bar m} e^{\sqrt{\frac{2}{k}} [ (m +
  \frac{k}{2} \frac{q}{N}) X +
  ( \bar m + \frac{k}{2} \frac{q}{N}) \bar X]}. \label{elop}
\end{equation}
For $q \in \mathbb{Z}_N$, these are ``fractional spectral flowed''
operators associated with the orbifold. For $q \in N \mathbb{Z}$,
these are the long string states in the global \ads3 covering space
\eqref{globalmet}. 

To take the flat space limit, we let $k \to \infty$, holding $x^3 =
\sqrt{k}\, t, r= \sqrt{k} \,\rho$ fixed. The metric becomes
\begin{equation}
ds^2 = -(dx^3)^2 + dr^2 + r^2 d\theta^2, 
\end{equation} 
so the orbifold goes over to the usual flat-space orbifold in this
limit. If we define $x^1 = r\,\cos \theta$, $x^2 = r \,\sin \theta$, we have,
 $J^a = i \sqrt{k} \,\partial x^a$ and $\bar J^a = i \sqrt{k} \,\bar \partial
x^a$. Since we hold $\theta$ and $\sqrt{k} \,t$ fixed, we should take
$Q_\theta = J^3_0 - \bar J^3_0 \sim 1$ and $Q_t = J^3_0 + \bar J^3_0
\sim \sqrt{k}$. This implies that 
\begin{equation}
m \to \frac{1}{2} (p_\theta + \sqrt{k} p_3 - k \frac{q}{N}), \qquad
\bar m \to \frac{1}{2} (-p_\theta + \sqrt{k} p_3 - k \frac{q}{N}).
\end{equation}
In the flat space limit, $X \approx -i\sqrt{2} \,x^3_L(z)$, $\bar X \approx
i\sqrt{2} \,x^3_R( \bar z)$, and the boson part of the operator
\eqref{elop} becomes just a momentum mode in the $x^3$ direction,
\begin{equation}
\Phi^q_{j m \bar m} \approx \Psi_{j m \bar m}\, e^{-ip_3 x^3}. 
\Label{ellop}
\end{equation} 
We also need to require that the parafermion part has a regular
limit. For the untwisted sectors, this requires $j \sim \sqrt{k}$. For
the twisted sectors, we need
\begin{equation} \label{jscalee}
j \to \frac{1}{2} (k \frac{q}{N} - \sqrt{k} p_3 + \alpha)  
\end{equation}
for some constant $\alpha$. However, recall that there is a bound on
$j$: we only allow current algebra representations with $\frac{1}{2}
< j < \frac{k-1}{2}$~\cite{Maldacena:2000hw}. This is consistent with
the required scaling~\eqref{jscalee} only for $q < N$. Hence, in the
flat space limit, the twisted sector operators for $q \in \ZZ_N$ have a
regular limit, and should give us the usual twisted sectors for the
flat space orbifold, but the long string states with $q \geq N$ go off
to infinite conformal dimension as we take the limit. Thus, we regain
precisely the expected flat space spectrum.

\section{Comments on the \ads3 partition function}
\label{partitionfn}

In this paper, we have approached the calculation of the twisted
sector spectrum on BTZ through a vertex operator construction. It
would be more satisfying to construct a modular-invariant partition
function for the string theory on BTZ and extract the spectrum for
this partition function. In fact, the appropriate partition function
has already been constructed for the bosonic string: Since Euclidean
BTZ is the same spacetime as thermal \ads3, the thermal \ads3
partition function calculated in~\cite{Maldacena:2000kv} can be
re-interpreted as a BTZ partition function. Alas, technical
difficulties have prevented us from extracting the spectrum from this
partition function. In this appendix, we will explain why this route
is obstructed. Identifying the tachyon from the Euclidean BTZ partition function
is also discussed in~\cite{Lin:2007gi,Micha}.

First, let us recall the relation between the two spacetimes. We have
the Euclidean metric
\begin{equation}
ds^2 = k\, (\cosh^2\!\! \rho \, d\tau^2 + d\rho^2  + \sinh^2 \!\!\rho \,
d\theta^2 )\ .
\Label{adsbtz}
\end{equation}	
Interpreting $\tau$ as time and $\theta$ as the spatial circle gives
us the global \ads3 metric; a further identification $\tau \sim \tau +
\beta$ gets us to thermal \ads3.  On the other hand we can take
$\theta$ to be the temporal direction and $\tau$ as the spatial
circle.  This is then the Euclidean BTZ black hole, as the temporal
circle shrinks smoothly to zero (at $\rho = 0$); a simple coordinate
change $\cosh \rho = r$ maps this back to usual BTZ coordinates
\req{btzmet}.  Formally, given a $\mathbf{T}^2$ parametrised by
$(\tau,\theta)$, we want a hyperbolic three-manifold which has the
$\mathbf{T}^2$ as its boundary. Which circle of the torus we choose to
make contractible in the bulk geometry determines the spacetime.  By
choosing to make a particular combination of the boundary one-cycles
contractible in the bulk we can construct a full $SL(2,\cC)$ family of
black holes ~\cite{Maldacena:1998bw}.  The different geometries are
related by an $SL(2,\cC)$ transformation of the boundary complex
structure. Note that this is a modular transformation from the point
of view of the dual boundary CFT. From the world-sheet point of view,
Euclidean BTZ and thermal \ads3 are the same spacetime, given
different interpretations corresponding to different ways in which we
can analytically continue to a Lorentzian spacetime.

Hence, the thermal AdS partition function calculated
in~\cite{Maldacena:2000kv} can also be interpreted as the BTZ
partition function. The contribution to the torus partition function
from the thermal AdS factor is, for a fixed world-sheet modular
parameter~\cite{Gawedzki:1991yu,Maldacena:2000kv} \footnote{In writing
  this expression we have corrected a typo in~\cite{Maldacena:2000kv}.
  We would like to thank James Lucietti for discussions on this
  issue.}
\begin{equation}
\CZ_{{\rm AdS}}(\b, \mu;\tau) ={ \beta \, \sqrt{k-2} \over 2 \pi \, \sqrt{\tau_2}}  \, \sum_{n,m} \, {e^{-k \b^2 |m-n\tau|^2/4\pi\tau_2 + 2\pi \, \Im(U_{n,m})^2/\tau_2 }  \over |\vth_1 (\tau, U_{n,m}) |^2 }
\Label{thermZ}
\end{equation}	
with 
\begin{equation}
U_{n,m}(\tau) = {i\over 2 \pi} \, \beta \, (1-i\mu) \, (n \bar{\tau} -m) \ . 
\Label{udef}
\end{equation}	
To obtain the full partition function, we need to add internal and
ghost contributions, and sum over the fundamental domain for the
world-sheet modular parameter. The sum over $n$ above can be traded for
a sum over copies of the fundamental domain, allowing us to write the
full partition function as
\begin{equation}
Z_{{\rm AdS}}(\beta,\mu) = 
\int_0^\infty \frac{d\tau_2}{\tau_2} \int_{-1/2}^{1/2} d\tau_1
e^{4\pi \tau_2 (1- \frac{1}{4(k-2)})} \sum_{h,\bar h} D(h,\bar h) q^h
\bar q^{\bar h} \CZ_{{\rm AdS}}(\b, \mu;\tau),
\end{equation}
where $D(h, \bar h)$ are the degeneracies in the internal CFT, and
$\CZ_{{\rm AdS}}(\b, \mu;\tau)$ should be understood as now only
involving a sum on $m$.  In~\cite{Maldacena:2000kv}, the spectrum on
\ads3 was extracted from this partition function by expanding in terms
of single string energy eigenstates.  To do so, we write the free
energy as
\begin{equation}
F(\b,\mu) = -{1\over \b} \, Z(\b ,\ \mu) 
= \frac{1}{\beta} \sum_{string \in \CH} \log \left( 1- e^{\beta \(
  E_{string} + i \mu \, \l_{string}\)} \right) = \sum_{m=1}^{\infty} \, f( m\beta, m \mu) \ ,
\Label{thermalAdSa}
\end{equation}
where 
\begin{equation}
f(\beta, \mu) = {1\over \beta} \sum_{string \in \CH} \, e^{ -\beta \(
  E_{string} + i \mu \, \l_{string}\)} 
\Label{thermalAdSb}
\end{equation}	
and $\CH$ is the single string Hilbert space. This would allow us to
read off the spectrum. The calculation is relatively straightforward,
as the sum on $m$ in \eqref{thermalAdSa} can be identified with the
sum on $m$ in \eqref{thermZ}.

To perform the same calculation in BTZ, we would want to write
\begin{equation} \label{BTZa}
F(\b_{BTZ},\mu_{BTZ}) = -{1\over \b_{BTZ}} \, Z(\b_{BTZ} ,\ \mu_{BTZ}) 
= \sum_{m'=1}^{\infty} \, f( m'\beta_{BTZ}, m' \mu_{BTZ}) \ ,
\end{equation}
where $\b_{BTZ}$ is the inverse temperature of the black hole, and
$f(\beta_{BTZ}, \mu_{BTZ})$ is as in \eqref{thermalAdSb}. The
problem with the calculation is that we can no longer identify the sum
on $m'$ in \eqref{BTZa} with the sum on $m$ in \eqref{thermZ}. Let us
consider for simplicity the case of zero chemical potential, $\mu =0$.
Then re-interpreting the thermal \ads3 space as the Euclidean BTZ
black hole gives $\beta_{BTZ} = 4\pi^2/\beta$, $\mu_{BTZ} = 0$. Thus,
the sum on $m'$ in \eqref{BTZa} is a sum in $m'/\beta$, whereas the
sum on $m$ in \eqref{thermZ} is a sum on $m \beta$. To extract the BTZ
spectrum, we need to rewrite \eqref{thermZ} in terms of a sum on
integer/$\beta$.

First, note that it is not possible to achieve the desired rewriting
by a world-sheet modular transformation. The two descriptions are
related by a modular transformation in the boundary theory, but not
from the world-sheet point of view. In the world-sheet, we are
considering the same Euclidean target space; we are only changing our
interpretation of it. 

What we need to do is to make a Poisson resummation over $(n,m)$ in
\eqref{thermZ}; this will replace the sum over $m \beta$ by a sum over
$p/\beta$ for an integer $p$. We have not been able to carry out this
Poisson resummation because of the theta function $\vth_1(\tau,
U_{n,m})$ appearing in the denominator of \req{thermZ}. One can expand
this factor in a power series, and perform the Poisson resummation
term by term, but it is then not possible to resum the power series to
obtain the desired information. Thus, although the partition function
in principle contains all the information we want, we have had to take
a vertex operator approach to identify the tachyons on the Lorentzian
BTZ black hole.

\bibliography{tachyonbtz}
\bibliographystyle{utphys}

\end{document}